\documentclass[journal,comsoc,twoside]{IEEEtran}

\usepackage[T1]{fontenc}
\usepackage{cite}
\usepackage[pdftex]{graphicx}
\graphicspath{{./figures/}}
\DeclareGraphicsExtensions{.pdf}
\usepackage{amsmath}
\usepackage[cmintegrals]{newtxmath}
\usepackage{bm}
\usepackage{trfsigns}
\usepackage{amssymb}
\usepackage{mathtools}
\usepackage{nicefrac}
\usepackage{algpseudocode}
\usepackage{framed}
\algrenewcommand{\algorithmiccomment}[1]{\hfill \textit{#1}}
\usepackage{setspace}
\usepackage{array}
\usepackage{booktabs}
\usepackage{url}
\usepackage{color}

\newcommand{\T}{\mathsf{T}}
\newcommand{\CT}{\mathsf{H}}
\newcommand{\norm}[1]{{\left\|#1\right\|}}
\newcommand{\abs}[1]{{\left|#1\right|}}

\newcommand{\C}{\mathbb{C}}

\newcommand{\eg}{e.g.}

\newcommand{\ie}{i.e.}

\newcommand{\SNR}{\textrm{SNR}}
\newcommand{\rank}{\operatorname{rank}}
\newcommand{\rect}{\operatorname{rect}}
\newcommand{\brackets}[1]{{\left(#1\right)}}
\newcommand{\rbrackets}[1]{{\left(#1\right)}}
\newcommand{\sbrackets}[1]{{\left[#1\right]}}
\newcommand{\cbrackets}[1]{{\left\{#1\right\}}}

\newcommand{\E}{\operatorname{E}}

\renewcommand{\vec}{\mathbf}
\renewcommand{\triangleq}{\stackrel{\text{\tiny def}}{=}}

\definecolor{red}{rgb}{0,0,0}

\begin{document}

\title{\color{red}An Adaptive CM Array Preconditioner for\\ Blind Multi-User Separation}

\author{Stanislaw~Gorlow, Jo\~{a}o~Paulo~C.~L.~da~Costa, and~Martin~Haardt%
\thanks{S. Gorlow is with Dolby Sweden, Stockholm,
Stockholm County, 113\,30 Sweden e-mail: szgorl@dolby.com. Parts of this work have been published in \cite{Gorlow2017_WCNPS}.}%
\thanks{J. P. C. L. da Costa is with the University of Bras\'{i}lia. This work was partially funded by the Coordena\c{c}{\~a}o de Aperfei\c{c}oamento de Pessoal de N\'ivel Superior (CAPES) under grant number 303905/2014-0.}%
\thanks{M. Haardt is with the Ilmenau University of Technology.}}%

\markboth{}{Gorlow \MakeLowercase{\textit{et al.}}: Adaptive CM Array Preconditioner}

\maketitle

\begin{abstract}
The family of constant-modulus algorithms is widely used in wireless communication systems and in radar. The classical constant-modulus adaptive (CMA) algorithm, however, fails to lock onto a single mode when used in conjunction with an antenna array. Instead, it equalizes the entire spatial spectrum. In this paper, we describe in full detail our recently proposed approach for the separation of multiple users in a radio system with frequency reuse, such as a cellular network, making use of the CMA algorithm. Based on the observation that the differential filter weights resemble a superposition of the array steering vectors, we cast the original task to a direction-of-arrival estimation problem. With rigorous theoretical analysis of the array response based on the discrete-space Fourier transform we elaborate a solution that solves the problem by finding the roots of a polynomial equation. We provide a numerical example to demonstrate the validity of the approach under high-SNR conditions. In addition, we propose a more general preprocessor for the CMA array which allows the modulated signals to differ in amplitude. As a byproduct, the preprocessor yields a low-cost estimate of the number of concurrent users, i.e.\ the model order, by simply counting the roots \textcolor{red}{with the strongest response}. 
\end{abstract}


\section{Introduction}

\IEEEPARstart{A}{}constant-modulus adaptive (CMA) algorithm represents a class of adaptive digital filtering algorithms that are sensitive to amplitude modulation and insensitive to angle modulation. Most commonly, it is realized as a transversal finite impulse response (FIR) digital filter based on gradient descent. In its original formulation the CMA algorithm compensates for the frequency-selective multipath and interference on signals that have a constant envelope. Such radio signals can be generated using frequency modulation or quadrature phase-shift keying (QPSK) \cite{Treichler1983_TASSP}. Despite its close resemblance to the least mean squares (LMS) filter, the main feature of the CMA filter is that it does not require a reference signal to operate. For a topic review, the reader is referred to \cite{Johnson1998}. 

In \cite{Treichler1983_TASSP} it is argued that the CMA filter is not only useful to counter multipath induced fading, but also to reduce narrow-band or sinusoidal interference if the interferer's power is significantly below the carrier power. In that case, the CMA filter acts as a notch filter. Nevertheless, in order to deal with a strong interferer, extra measures, such as prefiltering, are necessary. In \cite{Treichler1985_TASSP_Capture}, based on the model of two sinusoids at the input and nonoverlapping spectra, two different solutions can be found for the filter: one to suppress the interferer and another to capture the signal of interest. There are cases, however, in which the interferer is mistakenly captured as the signal of interest.

With regard to the convergence of the CMA array, we find works such as \cite{Shynk1993_ASILOMAR} or \cite{Keerthi1994_ASILOMAR}. Both rely on the capture effect of the CMA filter, \ie\ the assumption that the algorithm will lock onto the strongest CM signal with a particular polarization or direction of arrival. In \cite{Gooch1986_ICASSP}, however, it is mentioned that even in the case of two sources that are separated by an angle of 90\textsuperscript{$\circ$} the array's capture effect may fail, as the filter may switch arbitrarily from one source to the other. 
The problem of capturing one particular source in the presence of multiple CM sources is further acknowledged in \cite{Chen2004_TSP}. Without providing any evidence, it is inferred that the problem can be solved in multiple stages. For this to be possible, the CMA array would need to lock in each stage, which is not guaranteed. 

Various application areas for CM algorithms can be found in the literature. In \cite{Zhao2017}, a CM algorithm serves as the basis for semi-blind adaptive beamforming in an array-based communication system. In \cite{Balogun2017}, a CM algorithm is employed to estimate the carrier-frequency offset in an optical system. In \cite{Cheng2017}, a CM algorithm is used to design the probing waveform for a MIMO radar. In this paper, we present a complete derivation of our recent approach to the blind multi-user separation problem based on the CMA array \cite{Gorlow2017_WCNPS} and extend it by a more general adaptive preprocessor. The weights of the preprocessor are the coefficients of a polynomial in a single variable $z$ with the constant term equal to one. The number of roots of the polynomial that lie near the unit circle in the $z$-plane represent the model order, while the arguments of the complex roots correspond to the angles of the sources. From the angles one can construct a response matrix and \emph{condition} the CMA array in such a way that it is likely to focus on the user of interest and to reject all the competing users. 

The rest of the paper is organized as follows. In Section~\ref{sec:model_and_problem}, we define the data model of an antenna array and formulate the multi-user separation problem. In Section~\ref{sec:cma_algorithm}, we review the classical literature on the CMA algorithm and conclude the section with more remarks concerning the capture effect. Section~\ref{sec:dsft} defines the discrete-space Fourier transform that provides the mathematical basis for our approach. In Section~\ref{sec:root_cma_array}, we elaborate our extension of the CMA algorithm for the antenna array. \textcolor{red}{In Section~\ref{sec:preconditioner}, we present a more practical solution in the form of an adaptive preprocessor for the CMA array.} Conclusions are drawn in Section~\ref{sec:conclusion}. 

\section{Data Model and Problem Statement}
\label{sec:model_and_problem}

In this section, we present the data model for a frequency-reuse cellular network and state the blind multi-user separation problem which we seek to solve using the constant-modulus adaptive algorithm. We consider a scenario in which multiple sources (users) broadcast signals at the same frequency and at the same time at distinct locations from potentially different cells. The CM signals are received by a base station equipped with (structurally) identical antennas arranged as an array. By combining the antenna outputs, the objective is to separate the CM signals with minimal interference. 

\subsection{Data Model}

Consider the linear data model
\begin{equation}
\vec{X} = \vec{A} \, \vec{S}^\CT + \vec{N} \text{,}
\label{eq:model}
\end{equation}
where \textcolor{red}{$\vec{S}^\CT$ is the Hermitian transpose of the signal matrix $\vec{S} \in \C^{N \times D}$},
\begin{equation}
\vec{S} = {
	\begin{bmatrix}
		\vec{s}_1 & \vec{s}_2 & \cdots & \vec{s}_D
	\end{bmatrix}}
\end{equation}
with the $d$th column belonging to the $d$th source signal, $\vec{A} \in \C^{M \times D}$ is the array response matrix, the columns of which are vectors on the array manifold associated with a direction of arrival (DOA) or a spatial frequency,
\begin{equation}
\vec{A} = {
	\begin{bmatrix}
		\vec{a}_1 & \vec{a}_2 & \cdots & \vec{a}_D
	\end{bmatrix}} \text{,}
\end{equation}
$\vec{N} \in \C^{M \times N}$ is a (zero-mean) Gaussian noise matrix, and $\vec{X} \in \C^{M \times N}$ carries $N$ temporal snapshots collected by $M$ antennas ($N \gg M$). On the assumption that the antenna elements are arranged as a uniform linear array (ULA), $\vec{A}$ has the structure of a Vandermonde matrix, \ie
\begin{equation}
\vec{a}_d = {
	\begin{bmatrix}
		1 & z_d & z_d^2 & \cdots & z_d^{M - 1}
	\end{bmatrix}}^\T
	\quad \text{with}\ z_d = \e^{\im 2 \pi \xi_d} \text{,}
	\label{eq:vandermonde}
\end{equation}
where $\e$ is Euler's number, $\im$ is the imaginary unit, and $\xi_d$ is the spatial frequency, 
\begin{equation}
\xi_d = \frac{\Delta}{\lambda} \, \sin{\theta_d} \text{,} 
\end{equation}
where $\Delta$ is the spacing between the elements of the antenna array, $\lambda$ is the carrier wavelength, and $\theta_d$ is the inclination angle of the $d$th wavefront. We assume a stationary propagation environment in which the channel has a short delay spread, so that temporal equalization can be omitted. 

\subsection{Problem Statement}

In the noiseless case, the problem at hand can be formulated as a structured matrix factorization problem: 
\begin{equation}
\begin{aligned}
&\vec{X} = \hat{\vec{A}} \, \hat{\vec{S}}^\CT \\
&\text{s.t.}\ \rank{\hat{\vec{A}}} = \rank{\hat{\vec{S}}} = D\ \text{and}\ \abs{\hat{s}_{nd}} = 1 \text{.}
\end{aligned}
\end{equation}
The task is to find the factors $\hat{\vec{A}}$ and $\hat{\vec{S}}$ for a given $\vec{X}$, with $\hat{\vec{S}}$ satisfying the constant modulus property. In the beamforming context, the problem can also be formulated as: 
\begin{equation}
\begin{aligned}
&\vec{W}^\CT \, \vec{X} = \hat{\vec{S}}^\CT \\
&\text{s.t.}\ \rank \hat{\vec{S}} = D\ \text{and}\ \abs{\hat{s}_{nd}} = 1 \text{.}
\end{aligned}
\label{eq:output}
\end{equation}
Once $\vec{W}^\CT$ is known, an estimate of the array response matrix $\hat{\vec{A}}$ is given by the Moore--Penrose pseudoinverse of $\vec{W}^\CT$. As a consequence, $\vec{W}^\CT$ would provide the least-squares estimate for $\vec{S}$ under the constant modulus constraint. The task of the blind beamformer is to compute the proper weight vectors from the measured data only, without detailed knowledge of the signals and the channel. 

\section{Constant-Modulus Adaptive Algorithm}
\label{sec:cma_algorithm}

In this section, we give a review of the available literature related to the constant-modulus adaptive algorithm. We start from the single-channel transversal filter and end the section with the multi-channel spatial filter: the CMA array.

\subsection{Single-Channel Transversal Filter}

The original (single-channel) CMA filter is discussed at full length in \cite{Treichler1983_TASSP}. In the following, we give a summary of how to design it. First, consider a sampled quadrature signal $x \brackets{n}$ to be frequency modulated and multipath distorted. This signal, which is analytic and hence complex, passes though a tapped delay line with adjustable complex weights, which represents the transversal FIR filter with filter coefficients $\vec{w}$. Then, the complex filter output can be written as
\begin{equation}
y \brackets{n} = \vec{w}^\CT \brackets{n} \, \vec{x} \brackets{n} \text{,}
\label{eq:output}
\end{equation}
where
\begin{equation}
\vec{x} \brackets{n} = {\begin{bmatrix}
	x \brackets{n} & x \brackets{n - 1} & \cdots & x \brackets{n - N + 1}
\end{bmatrix}}^\T
\label{eq:signal}
\end{equation}
is the data in the delay line at sampling instant $n$ and
\begin{equation}
\vec{w} \brackets{n} = {\begin{bmatrix}
	w_0 \brackets{n} & w_1 \brackets{n} & \cdots & w_{N - 1} \brackets{n}
\end{bmatrix}}^\T
\end{equation}
is the vector of $N$ adjustable filter coefficients. Conveniently, we assume that the coefficients are tweaked for each $n$.

\begin{figure}[!t]
\centering
\includegraphics[height=.9\columnwidth]{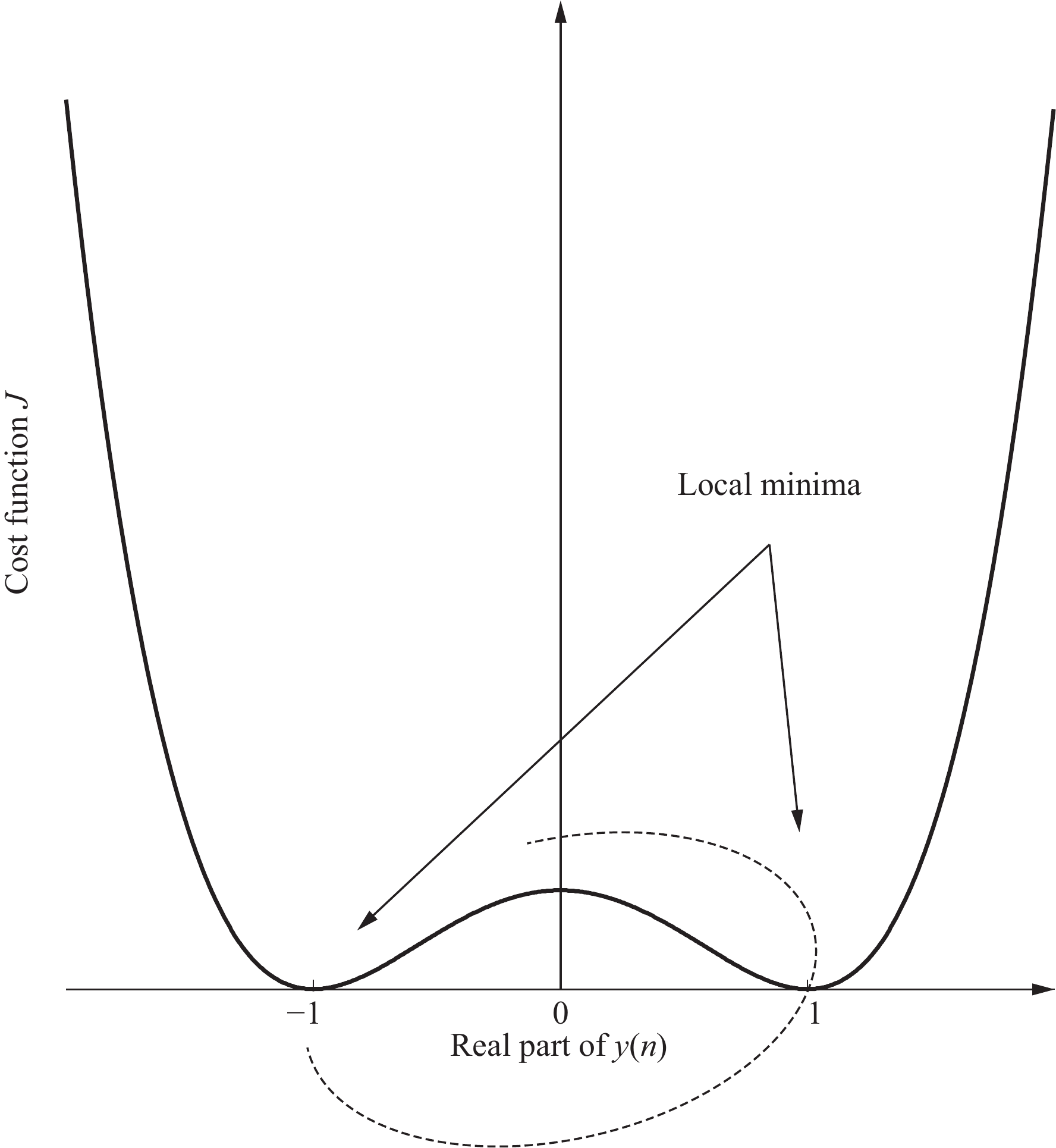}
\caption{$J$ as a function of $\Re{\cbrackets{y\brackets{n}}}$ for $\Im{\cbrackets{y\brackets{n}}} = 0$. Any local minimum on the unit circle around the origin satisfies $J = 0$. Positive deviations from unit modulus are penalized more than negative deviations.}
\label{fig:performance}
\end{figure}

The objective is to restore the output $y \brackets{n}$ to a form, which on average has a constant instantaneous modulus. It is achieved by choosing the coefficients $\vec{w}$ in such a way that the following cost or performance function is minimized:
\begin{equation}
J = \frac{1}{4} \E{\cbrackets{\sbrackets{\abs{y \brackets{n}}^2 - 1}^2}} \text{,}
\label{eq:cost_function}
\end{equation}
where $\E$ denotes expectation. Note that in \eqref{eq:cost_function} the modulus of the unknown transfer signal is normalized to unity. Hence, $J$ is a differentiable positive measure of the average amount by which the output $y \brackets{n}$ deviates from unit modulus, see Fig.~\ref{fig:performance}. Also note the inherent phase ambiguity in the cost.

With the cost function and the filter structure given, \eqref{eq:cost_function} is minimized using stochastic gradient descent. In particular, $\vec{w}$ is updated according to the recurrence relation
\begin{equation}
\vec{w} \brackets{n + 1} = \vec{w} \brackets{n} - \gamma \, \nabla_\vec{w}{J\brackets{n}} \text{,}
\label{eq:recurrence_relation}
\end{equation}
where $\gamma$ is the step size (adaptation constant) and $\nabla_\vec{w}$ is the gradient with respect to the filter coefficients. The stochastic gradient, 
\begin{equation}
\nabla_\vec{w}{J\brackets{n}} = \E{\cbrackets{\sbrackets{\abs{y \brackets{n}}^2 - 1} \, \vec{x} \brackets{n} \, y^\ast \brackets{n}}} \text{,} 
\label{eq:stochastic_gradient}
\end{equation}
can be approximated by the instantaneous gradient leaving out the expectation. The update rule from \eqref{eq:recurrence_relation} then reads:
\begin{equation}
\vec{w} \brackets{n + 1} = \vec{w} \brackets{n} - \gamma \, \sbrackets{\abs{y \brackets{n}}^2 - 1} \, \vec{x} \brackets{n} \, y^\ast \brackets{n} \text{.}
\label{eq:update_rule}
\end{equation}
By defining
\begin{equation}
\sbrackets{1 - \abs{y \brackets{n}}^2} \, y \brackets{n} \triangleq e \brackets{n} \text{,}
\label{eq:cma_error}
\end{equation}
the update rule \eqref{eq:update_rule} becomes
\begin{equation}
\vec{w} \brackets{n + 1} = \vec{w} \brackets{n} + \gamma \, \vec{x} \brackets{n} \, e^\ast \brackets{n} \text{,}
\label{eq:lms_filter}
\end{equation}
which appears to be formally identical with the LMS filter. 

\color{red}
Please note that $y(n) = 0$ is also a stationary point (a local maximum) for the cost function $J$, see \eqref{eq:stochastic_gradient}. However, due to \eqref{eq:output} and since $\vec{x}(n) \neq \vec{0} \ \forall \, n$, this point can only be reached iff $\vec{w}(n) = \vec{0}$. This is avoided by setting $\vec{w} \neq \vec{0}$ at initialization. 
\color{black}

\subsection{Multi-Channel Transversal Filter}

In \cite{Treichler1985_TASSP}, a multi-channel extension of the above CMA filter is presented in the form of a polarization diversity combiner. By means of polarization beam steering, the combiner rejects additive interference and compensates for channel-induced polarization rotation at the same time. The authors provide an example in which a two-channel processor with one channel for the vertical and the other for the horizontal polarization successfully separates an FM signal from a QPSK signal, both transmitted over orthogonal channels that exhibit frequency-selective multipath and cross talk. It should be noted that the two signals have an identical nominal carrier frequency and the same power. And although the result is obtained in the high-SNR regime, \ie, at an SNR of 20 dB at the receiver feeds, it looks promising. According to the authors, the multi-channel processor can be employed in combination with any other orthogonal scheme. Still, it is not quite clear what such a scheme would look like for more than two transmitter feeds. Two possible implementations of a two-channel processor are discussed below.

\subsubsection{One-Stage Processor}

The one-stage processor has one pair of weight vectors, $\vec{w}_1$ and $\vec{w}_0$, for a CM signal transmitted on either the horizontal or the vertical feed. Each tapped delay line on the receiver side is initialized in such a way as to have an all-pass response for the desired signal or to fully attenuate the cross polarization by setting all weights to zero. The output from the two filters is summed to yield the desired CM signal by which the weight vectors are adjusted. Hence, if $x_h \brackets{n}$ is the signal of interest and $x_v \brackets{n}$ is the interferer, the estimate is formed by
\begin{equation}
y_h \brackets{n} = \vec{w}_1^\CT \brackets{n} \, \vec{x}_h \brackets{n} + \vec{w}_0^\CT \brackets{n} \, \vec{x}_v \brackets{n} \text{,}
\label{eq:horizontal}
\end{equation}
where
\begin{equation}
\begin{aligned}
\vec{w}_1 \brackets{n_0} &= {\begin{bmatrix} 1 & 0 & \cdots & 0 \end{bmatrix}}^\T \text{,} \\
\vec{w}_0 \brackets{n_0} &= {\begin{bmatrix} 0 & 0 & \cdots & 0 \end{bmatrix}}^\T \text{,}
\end{aligned}
\end{equation}
and
\begin{equation}
\begin{aligned}
\vec{x}_h \brackets{n} &= {\begin{bmatrix}
	x_h \brackets{n} & \cdots & x_h \brackets{n - N + 1}
\end{bmatrix}}^\T \text{,} \\
\vec{x}_v \brackets{n} &= {\begin{bmatrix}
	x_v \brackets{n} & \cdots & x_v \brackets{n - N + 1}
\end{bmatrix}}^\T \text{.}
\end{aligned}
\end{equation}
By stacking the weight vectors to a single vector
\begin{equation}
\vec{w}_h \brackets{n} = {\begin{bmatrix} \vec{w}_1 \brackets{n} \\ \vec{w}_0 \brackets{n} \end{bmatrix}}
\end{equation}
and by doing the same for the signal vectors,
\begin{equation}
\vec{x} \brackets{n} = {\begin{bmatrix} \vec{x}_h \brackets{n} \\ \vec{x}_v \brackets{n} \end{bmatrix}} \text{,}
\end{equation}
we see that \eqref{eq:horizontal} is equivalent to
\begin{equation}
y_h \brackets{n} = \vec{w}_h^\CT \brackets{n} \, \vec{x} \brackets{n} \text{,}
\label{eq:horizontal_short}
\end{equation}
which again is identical with the single-channel filter \eqref{eq:output}. As a result, we may state that space diversity is not exploited in that scheme. The interference rejection performance is hence subject to the degree of orthogonality of the transmitter feeds, the mechanical alignment of the transmitter with the receiver feeds, the channel-induced polarization rotation, but also the SIR at the receiver. As a rule of thumb, the stronger the CM signal and the weaker the cross talk, the better the outcome. With the above definitions, the processor is updated as per
\begin{equation}
\begin{aligned}
{\begin{bmatrix}
	\vec{w}_h \brackets{n + 1} \\
	\vec{w}_v \brackets{n + 1}
\end{bmatrix}} &= {\begin{bmatrix}
	\vec{w}_h \brackets{n} \\
	\vec{w}_v \brackets{n}
\end{bmatrix}} - \gamma \, \cbrackets{{\begin{bmatrix}
	\abs{y_h \brackets{n}}^2 \\
	\abs{y_v \brackets{n}}^2
\end{bmatrix}} - {\begin{bmatrix} 1 \\ 1 \end{bmatrix}}} \\
&\qquad {} \cdot \vec{x} \brackets{n} \, {\begin{bmatrix}
	y_h^\ast \brackets{n} \\
	y_v^\ast \brackets{n}
\end{bmatrix}}
\end{aligned}
\label{eq:single_stage}
\end{equation}
with
\begin{equation}
\vec{w}_v \brackets{n} = {\begin{bmatrix}
	\vec{w}_0 \brackets{n} \\ \vec{w}_1 \brackets{n}
\end{bmatrix}}
\end{equation}
\textcolor{red}{
and 
\begin{equation}
y_v \brackets{n} = \vec{w}_v^\CT \brackets{n} \, \vec{x} \brackets{n} \text{,}
\end{equation}
respectively}. 

\subsubsection{Two-Stage Processor}

The above single-stage processor can be realized in two stages with an adaptive signal canceller (ASC) in between. The approach has shown to be more robust to alignment imperfections and polarization crosstalk between two equally powered signals. The first stage filters out the FM signal, followed by the QPSK signal in the second stage. This order is preferred because the CMA filter has a more negative effect on the QPSK signal in the presence of an interferer. As a direct consequence, the SIR of the QPSK signal improves after removing the FM signal from the receiver inputs using the LMS filter. The error $e \brackets{n}$ in that case is defined as
\begin{equation}
e \brackets{n} \triangleq y \brackets{n} - \underbrace{\textcolor{red}{\vec{u}}^\CT \brackets{n} \, \vec{x} \brackets{n}}_{\hat{y} \brackets{n}} \text{,}
\end{equation}
where \textcolor{red}{$\vec{u}$} denotes the LMS filter coefficients. The input to the second stage is given by
\begin{equation}
\hat{\vec{x}} = {\begin{bmatrix}
	\hat{x}_h \brackets{n} \\
	\hat{x}_v \brackets{n}
\end{bmatrix}} = {\begin{bmatrix}
	x_h \brackets{n} \\
	x_v \brackets{n}
\end{bmatrix}} - {\begin{bmatrix}
	\textcolor{red}{\vec{u}}_1^\CT \brackets{n} \, \vec{x}_h \brackets{n} \\
	\textcolor{red}{\vec{u}}_0^\CT \brackets{n} \, \vec{x}_v \brackets{n}
\end{bmatrix}} \text{,}
\end{equation}
and hence
\begin{equation}
y_v \brackets{n} = \vec{w}_v^\CT \brackets{n} \, \hat{\vec{x}} \brackets{n} \text{.}
\end{equation}
Theoretically, this scheme can be extended to any number of stages. To put it into practice, all that is needed is one CMA filter and another LMS filter combined in recursion \cite{Treichler1985_TASSP}.

\subsection{Multi-Channel Spatial Filter}

A multi-channel spatial (CMA) filter is easily derived from \eqref{eq:output}, \eqref{eq:lms_filter}, and \eqref{eq:cma_error}. Changing the independent time variable $n$ in \eqref{eq:signal} to the channel index $m$ indicating an antenna element, we instantly obtain the formulation of a CMA array that can be put into effect in an adaptive antenna framework \cite{Gooch1986_ICASSP}. One would then refer to the CMA filter as a ``beamformer'' that steers the beam in the direction of the CM signal and rejects any additive interference. The array's weights thus determine the ``beam pattern''. The adaptation of the weights rests upon the fact that interference causes fluctuations to a CM signal's amplitude. The CMA array is meant, or at least expected, to counteract such fluctuations \cite{Treichler1983_TASSP}.

In the presence of several CM signals, the CMA array can also be implemented as a multi-stage processor, as shown in the previous subsection, to improve the SIR for the follow-up stage. In addition, its \textcolor{red}{convergence} performance can considerably be improved by adaptation of the step size $\gamma$ in reference to the recursive least squares (RLS) filter \cite{Plackett1950,Gooch1986_ICASSP,Chen2004_TSP}:
\begin{equation}
\gamma_n = \frac{\vec{P} \brackets{n - 1}}{\alpha + \vec{x}^\CT \brackets{n} \, \vec{P} \brackets{n - 1} \, \vec{x} \brackets{n}}
\label{eq:orthogonal}
\end{equation}
with
\begin{equation}
\begin{aligned}
\vec{P} \brackets{n} &= \sbrackets{1 - \frac{\vec{P} \brackets{n - 1} \, \vec{x} \brackets{n} \, \vec{x}^\CT \brackets{n}}{\alpha + \vec{x}^\CT \brackets{n} \, \vec{P} \brackets{n - 1} \, \vec{x} \brackets{n}}} \\
&\qquad {} \cdot \frac{1}{\alpha} \, \vec{P} \brackets{n - 1} \text{,}
\end{aligned}
\end{equation}
where $\alpha$ is the forgetting factor, usually chosen between $0.98$ and $1$, and $\vec{P} \brackets{n}$ denotes the inverse of the weighted sample correlation matrix for $\vec{x} \brackets{n}$. The latter is initialized as
\begin{equation}
\vec{R}_{\vec{x} \vec{x}} \brackets{n_0} = \sigma^2 \, \vec{I}_M \text{,}
\end{equation}
where $\vec{I}_M$ is an $M$-by-$M$ identity matrix and $\sigma$ has a positive value. Note that in \eqref{eq:orthogonal} $\gamma_n$ is no longer a scalar but rather a normalized inverse of $\vec{R}_{\vec{x}\vec{x}} \brackets{n}$. That being the case, $\gamma_n$ has the function of orthogonalizing the modes of $\vec{x} \brackets{n}$, which in turn speeds up the convergence.

\subsection{Further Remarks}

\begin{figure*}[!t]
\centering
\includegraphics[height=.9\columnwidth]{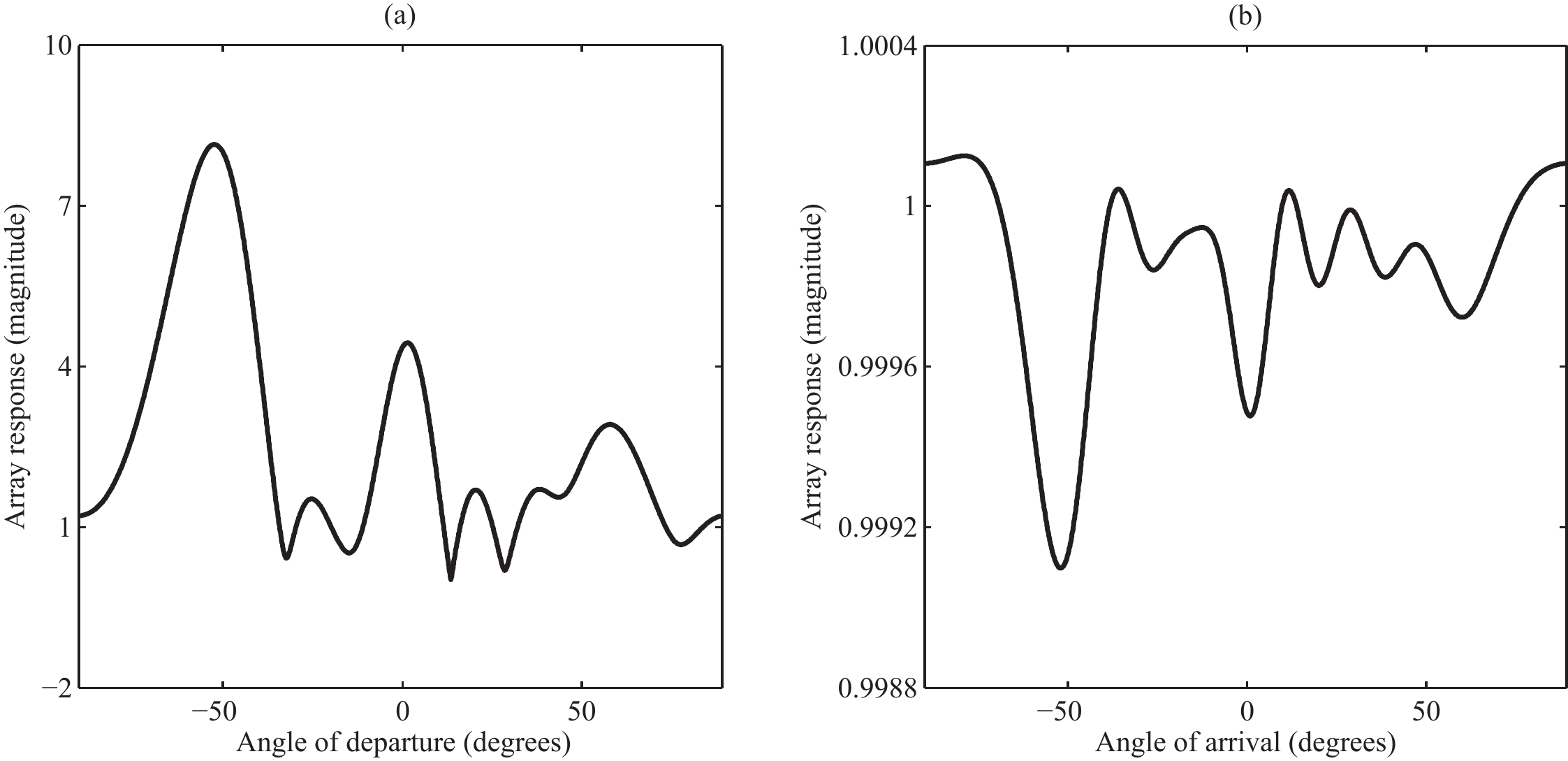}
\caption{Illustrative example of (a) the radiation pattern of three QPSK sources and (b) the response of the CMA array \textcolor{red}{(baseband)}.}
\label{fig:responses}
\end{figure*}

In \cite{Treichler1985_TASSP_Capture}, in the attempt to predict the algorithm's behavior, the authors derive conditions that allow to control the algorithm's general tendency. They conclude that the CMA filter's ability to capture the desired signal is subject to a combination of the following factors: the signals' amplitudes at the input and the output, \ie\ the SIRs, and the initial filter response. It should be pointed out, however, that most of the analysis applies only to the special case where the two time series are, in a broader sense, \emph{orthogonal} (see polarization). Furthermore, the authors also mention that according to their convergence analysis it is not sufficient to initialize the filter with an all-pass response when the amplitude of the signal of interest is near unity. The amplitude of the interferer has to be considered as well.

Especially from a spatial filtering perspective, where space diversity and frequency reuse play a major role, it is hard to see from the cost function why the phased array should steer the beam in the direction of one CM source but drive to zero the contribution from all other CM sources. Firstly, all QPSK signals impinging on the array \textcolor{red}{could} be considered as equally powered due to the far-field assumption. Secondly, since no reference is given in \eqref{eq:cost_function}, mutual independence of the set of CM sources is not exploited and so is irrelevant. A complete suppression of the interfering CM sources is realistic only if nulls are put explicitly in their directions by means of linear constraints, or by computation of the pseudoinverse of $\vec{A}$. In both cases, the angles of the incoming wavefronts have to be known a priori. This, however, is an unrealistic scenario. 

In fact, our experiments suggest that the CMA array more likely processes the CM sources as a whole, in proportion to the sources' amplitudes. Therefore, it does not seem to show any preference towards one particular CM source, neither for the strongest. The totality of all CM sources is treated as one single source with several multipath components, in the case of a channel with a long delay spread, or as several but fully correlated sources, if the delay spread is short. For this, it is impossible for the CMA array to distinguish several users in a frequency-reuse radio system that uses, \eg, QPSK. What it does instead is to adjust the filter weights in such a way that the filtered sum of all CM signals exhibits unit modulus. By transmitting two signals on different polarizations, the SIR is improved due to crosstalk reduction, given that the feeds are mechanically aligned. For all the reasons adduced above, we draw the conclusion that there is no capture effect that could be assigned to the CMA filter in a general sense. 

In summary it can be stated that the CMA array initialized with an all-pass response does not capture one particular CM signal within a mixture, but rather whitens the mixture. That means that the CMA array equalizes the channel inclusive of the radiation pattern of a phased array. This basic explanation is furthermore consistent with every observation or statement made by other authors. Fig.~\ref{fig:responses} illustrates the case where three QPSK sources send data from different angles and at different power levels. The corresponding radiation pattern is shown in (a). The response of the CMA array after 8,000 iterations is shown in (b), averaged over 100 simulations. It is worth to be pointed out that the algorithm did not converge. The average output modulus is $\E \sbrackets{\abs{y \brackets{n}}} = 1.084$. The CMA array acts as a soft equalizer of the radiation pattern without locking onto any mode. The array consists of 8 elements and the SNR is 20 dB with respect to the strongest mode.

\section{Discrete-Space Fourier Transform}
\label{sec:dsft}

Before solving the multi-user separation problem using the output from the original single-stage CMA array, let us introduce the discrete-space Fourier transform. The latter, together with the $z$-transform, is made use of in the following Section \ref{sec:root_cma_array}, so as to have a formal basis for the pursued approach and to justify the results.

\subsection{Definition}

The discrete-space Fourier transform (DSFT) of a series of real (or complex) numbers is a Fourier series that is periodic w.r.t.\ the angular spatial frequency variable $\mu$, also known as the angular wave number. The DSFT is defined as:
\begin{equation}
x\rbrackets{m}\ \fourier\ X{\left(\e^{\im \, \mu}\right)} = \sum_{m = -\infty}^\infty x(m) \, \e^{-\im \, \mu m} \text{.}
\label{eq:dsft}
\end{equation}
The corresponding inverse is given by
\begin{equation}
X\rbrackets{\e^{\im \, \mu}}\ \Fourier\ x(m) = \frac{1}{2 \pi} \int_{2 \pi} X{\left(\e^{\im \, \mu}\right)} \, \e^{\im \, \mu m} \, d \mu \text{.}
\end{equation}

\subsection{Transform Pairs and Properties}

Although many more transform pairs may be found in the vast amount of literature on the Fourier transform, let us seek out the two that are most relevant to our problem:
\begin{table}[!h]
\caption{Most Relevant Transform Pairs}
\centering
\begin{tabular}{*{3}{c}}
\toprule
Space series & & Fourier transform \\
$x(m)$ & \fourier & $X{\left(\e^{\im \, \mu}\right)}$ \\
\midrule
1 ($-\infty < m < \infty$) & & $\displaystyle 2 \pi \sum_{\kappa = -\infty}^\infty \delta {\left(\mu - 2 \pi \kappa\right)}$ \\
$x(m) = \begin{dcases*} 1 & for $0 \leqslant m < M$ \\ 0 & otherwise \end{dcases*}$ & & $\displaystyle \e^{-\im \, \mu \frac{M - 1}{2}} \frac{\sin{\left(M \frac{\mu}{2}\right)}}{\sin{\left(\frac{\mu}{2}\right)}}$ \\
\bottomrule
\end{tabular}
\label{tab:transform_pairs}
\end{table}

\noindent Note that the Fourier transform of the second pair is a variant of the Dirichlet kernel for a sum over all nonnegative $m$. The corresponding space series is equivalent in notation to
\begin{equation}
x(m) = \rect{\left[\frac{m - \frac{M - 1}{2}}{M - 1}\right]} \text{,}
\label{eq:rectangle}
\end{equation}
where $\rect{\cdot}$ denotes the rectangular function. In Table~\ref{tab:transform_pairs}, $\delta{(\mu)}$ stands for the Dirac delta function. To complete this section, let us recall two main operations in the space domain and the effect they have on the frequency domain.

\subsubsection{Shift in Frequency}

The first property is also referred to as frequency modulation by a (real) offset $\mu_0$. The respective Fourier transform pair is:
\begin{equation}
\e^{\im \, \mu_0 m} \, x(m)\ \fourier\ X{\left[\e^{\im \, {\left(\mu - \mu_0\right)}}\right]} \text{.}
\label{eq:modulation}
\end{equation}

\subsubsection{Multiplication in Space}

The other property states that a multiplication of two sequences in the space domain results in a periodic convolution in the frequency domain. Thus, the corresponding transform pair is:
\begin{equation}
w(m) \, x(m)\ \fourier\ \frac{1}{2 \pi} \int_{-\pi}^\pi W{\left(\e^{\im \, \nu}\right)} \, X{\left[\e^{\im \, {\left(\mu - \nu\right)}}\right]} \, d \nu \text{.}
\label{eq:multiplication}
\end{equation}

\noindent In the following subsections, we will use the transform pairs and properties from above to derive an analytic expression of the response of a beamformer such as the CMA array.

\subsection{Transform of a Finite-Length Sequence}

The expression in \eqref{eq:dsft} applies to an infinite series $x(m)$. In order to analytically evaluate the DSFT of a finite-length data sequence, we apply a rectangular window $w(m)$ of length $M$ to the input sequence $x(m)$, resulting in
\begin{align}
X{\left(\e^{\im \, \mu}\right)} &= \sum_{m = -\infty}^\infty w(m) \, x(m) \, \e^{-\im \, \mu m} \label{eq:windowed_series} \\
&= \sum_{m = 0}^{M - 1} x(m) \, \e^{-\im \, \mu m} \label{eq:finite_series}
\end{align}
with $w(m)$ as in \eqref{eq:rectangle}. Using the transform pair from \eqref{eq:multiplication}, we can see that the DSFT $X{\left(\e^{\im \, \mu}\right)}$ of a finite-length sequence as in \eqref{eq:finite_series} is formally identical with
\begin{equation}
\begin{aligned}
\widetilde{X}{\left(\e^{\im \, \mu}\right)} &= \frac{1}{2 \pi} \int_{-\pi}^\pi \e^{-\im \, \nu \frac{M - 1}{2}} \frac{\sin{\left(M \frac{\nu}{2}\right)}}{\sin{\left(\frac{\nu}{2}\right)}} \\
&\qquad {} \cdot X{\left[\e^{\im \, {\left(\mu - \nu\right)}}\right]} \, d \nu \text{.}
\end{aligned}
\label{eq:approximation}
\end{equation}
In other words, to compute the DSFT of a \emph{finite} series is the same as to convolve the transform of an \emph{infinite} series with the causal Dirichlet kernel, see Table~\ref{tab:transform_pairs}. Likewise, \eqref{eq:approximation} is a Fourier  series approximation of degree $M - 1$, see \cite{Grandke1983,Zhang2001}. 

\subsection{Fourier Series and Array Response}

\begin{figure}[!t]
\centering
\includegraphics[height=.9\columnwidth]{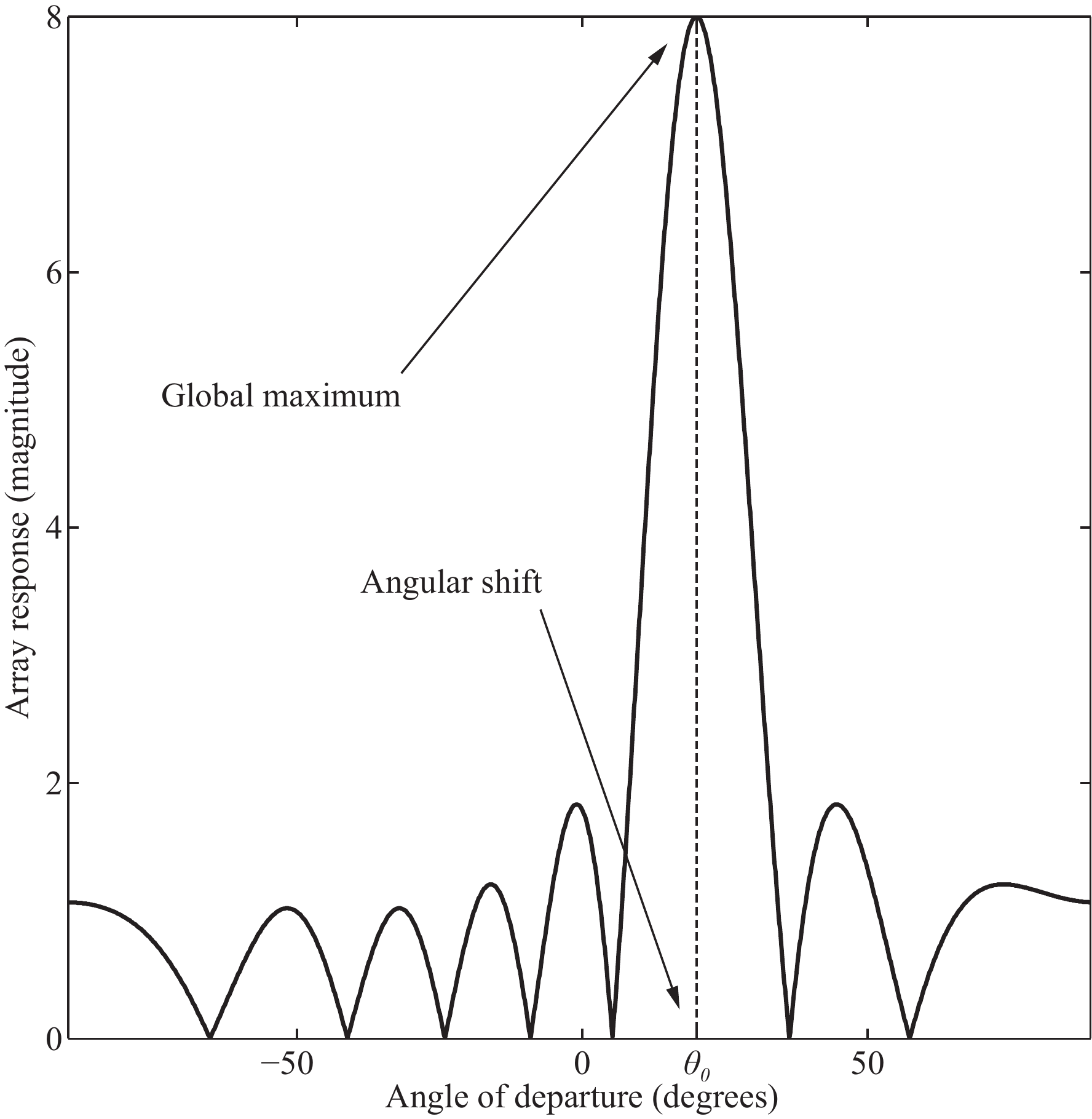}
\caption{Magnitude response of a steering vector ($M = 8$, \textcolor{red}{$\Delta = \nicefrac{\lambda}{2}$,} $\theta_0 = 20^\circ$).}
\label{fig:array_transform}
\end{figure}

Consider the case where we seek to compute the DSFT of the array steering vector consisting of the elements $a_m = \e^{\im \, \mu_0 m}$, $m = 0, 1, \dots, M - 1$, which can be written as $a(m) = \e^{\im \, \mu_0 m} \, x(m)$ with $x(m) = 1$. Resorting to the first correspondence from Table~\ref{tab:transform_pairs} and due to \eqref{eq:modulation}, once can easily show that, for $-\infty \leqslant m \leqslant \infty$,
\begin{equation}
a(m)\ \fourier\ A{\left(\e^{\im \, \mu}\right)} = 2 \pi \sum_{\kappa = -\infty}^\infty \delta {\left(\mu - \mu_0 - \ 2 \pi \kappa\right)} \text{.}
\end{equation}
In further consequence, for $m = 0, 1, \dots, M - 1$,
\begin{align}
A{\left(\e^{\im \, \mu}\right)} &= \frac{1}{2 \pi} \int_{-\pi}^\pi \e^{-\im \, {\left(\mu - \nu\right)} \frac{M - 1}{2}} \, \frac{\sin{\left(M \frac{\mu - \nu}{2}\right)}}{\sin{\left(\frac{\mu - \nu}{2}\right)}} \nonumber \\
&\qquad {} \cdot 2 \pi \, \delta {\left(\nu - \mu_0\right)} \, d \nu \nonumber \\
&= \e^{-\im \, {\left(\mu - \mu_0\right)} \frac{M - 1}{2}} \, \frac{\sin{\left(M \frac{\mu - \mu_0}{2}\right)}}{\sin{\left(\frac{\mu - \mu_0}{2}\right)}} \label{eq:array_transform} \text{.}
\end{align}
To derive \eqref{eq:array_transform}, we use the sifting property of the Dirac delta function. From \eqref{eq:array_transform} we hence conclude that the DSFT of the array steering vector is given by the (causal) Dirichlet kernel shifted by $\mu_0$ from zero. It has a global maximum at $\mu_0$ with a value of
\begin{equation}
A_{\max}{\left(\e^{\im \, \mu}\right)} = \lim_{\mu \rightarrow \mu_0} A{\left(\e^{\im \, \mu}\right)} = M \text{,}
\label{eq:maximum}
\end{equation}
see Fig.~\ref{fig:array_transform}.

The relation to the array response is established by taking a closer look at its definition. It is very common to write the output as in \eqref{eq:output}. Hence, the array or beam response $b{\left(\mu\right)}$ can be formulated as
\begin{equation}
b \brackets{\mu} = \vec{w}^\CT \, \vec{a} \brackets{\mu} \text{,}
\label{eq:beam_response}
\end{equation}
which is a function of the direction and the weights. Now, if we plug the elements of the steering vector $a_m\rbrackets{\mu}$ into \eqref{eq:beam_response}, it reads
\begin{equation}
b \brackets{\mu} = \sum_{m = 0}^{M - 1} w_m^\ast \, \e^{\im \, \mu m} \text{.}
\label{eq:unfolded_beam_response}
\end{equation}
Comparing \eqref{eq:unfolded_beam_response} with \eqref{eq:finite_series}, we see that
\begin{equation}
b \brackets{\mu} \equiv W^\ast \brackets{\e^{\im \, \mu}} \text{.}
\end{equation}
The equation above shows the identity between the response of a ULA, which is described by the steering vector, and the complex conjugate of the DSFT of the weights. Thus, one can conclude that the magnitude responses are identical, whereas the phase responses are mirrored about the $\mu$-axis. From this, one can \textcolor{red}{deduce} that the beam pattern, which is the magnitude squared of the beam response, is identical with the DSFT of the weights, when taking the magnitude squared of it, too. A final remark is that \eqref{eq:maximum} is equivalent to evaluating the beam response at $\mu_0$ with the weights adjusted according to $w_m = \e^{\im \, \mu_0 m}$, which is the same as computing the squared $\ell_2$-norm of $\vec{a}$, since
\begin{equation}
\norm{\vec{a}}_2^2 = \vec{a}^\CT \, \vec{a} = \sum_{m = 0}^{M - 1} \abs{a_m}^2 = M \text{.}
\label{eq:vectornorm}
\end{equation}
The squared $\ell_2$-norm of the steering vector is thus given by the number of the array elements $M$. Also note that the DSFT derived in this section refers to an array response matrix that has the Vandermonde structure. This further implies that the first array element is the reference element, which is lying in the origin of the space domain.

\subsection{Transform of a Sum of Steering Vectors}

\begin{figure}[!t]
\centering
\includegraphics[height=.9\columnwidth]{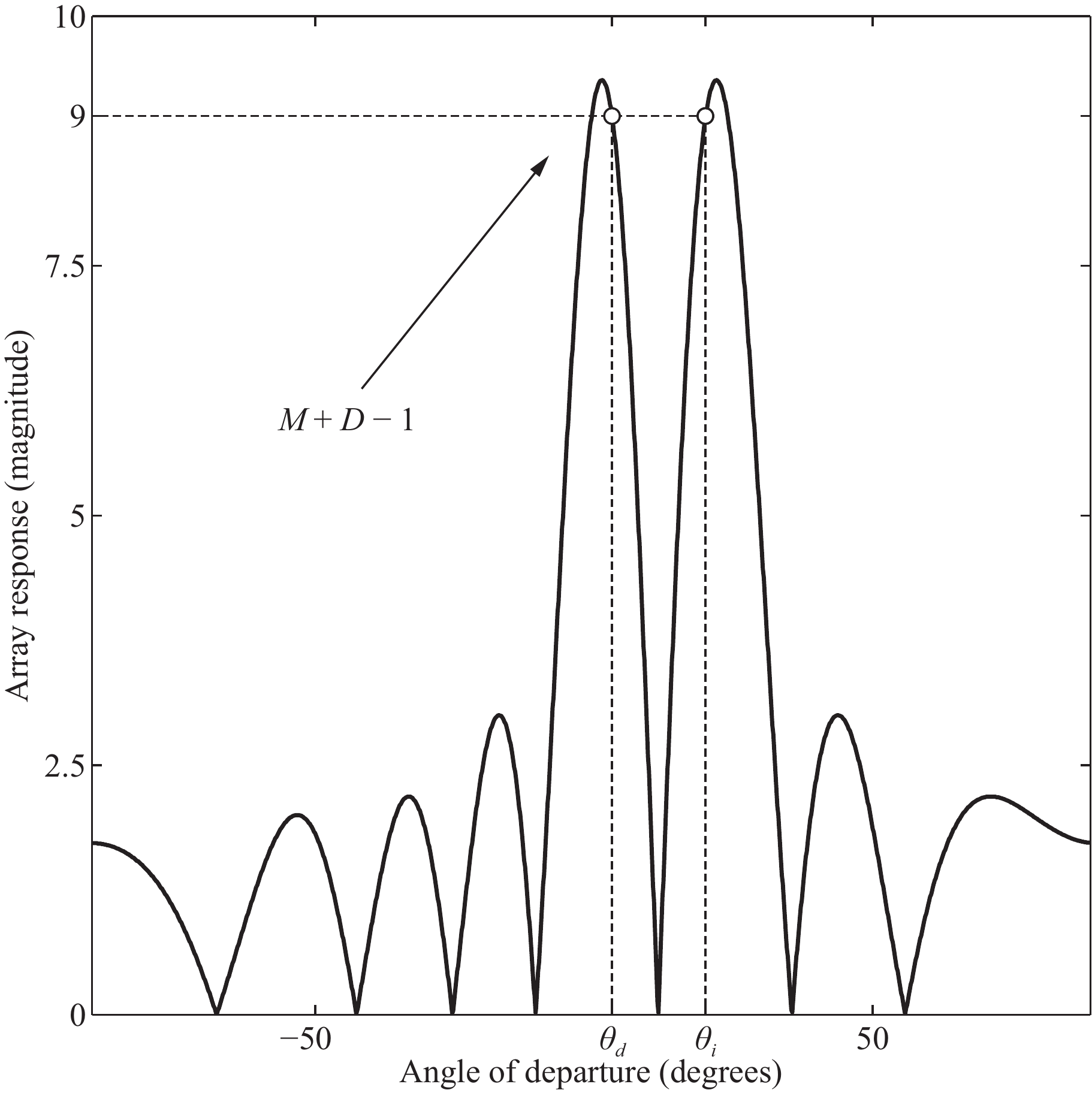}
\caption{Magnitude response of two superposing steering vectors for the case where $\Delta \mu = \nicefrac{2 \pi}{M - 1}$ ($M = 8$, \textcolor{red}{$\Delta = \nicefrac{\lambda}{2}$,} $\theta_d = 3.23^\circ$, $\theta_i = 20.0^\circ$).}
\label{fig:phase_relation}
\end{figure}

We close this part with a brief analysis of the special case where the space sequence is generated by superposition of a number of distinct array vectors, \ie,
\begin{equation}
a \brackets{m} = \sum_{d = 1}^D \vec{a}_{d_m} = \sum_{d = 1}^D \e^{\im \, \mu_d m} \text{,}
\label{eq:vecor_sum}
\end{equation}
for $m = 0, 1, \dots, M- 1$. Since the DSFT is linear, it follows that the DSFT of a sum of finite space sequences  is equal to the sum of DSFTs of the sequences alone. And therefore,
\begin{equation}
A \brackets{\e^{\im \, \mu}} = \sum_{d = 1}^D \e^{\im \, \brackets{\mu - \mu_d} \frac{M - 1}{2}} \, \frac{\sin \brackets{M \frac{\mu - \mu_d}{2}}}{\sin \brackets{\frac{\mu - \mu_d}{2}}}
\end{equation}
with
\begin{equation}
\begin{aligned}
&\left. A \brackets{\e^{\im \, \mu}}\right|_{\mu = \mu_i} \overset{\eqref{eq:maximum}}{=} M \\ 
&\qquad {} + \sum_{\substack{d = 1 \\ d \neq i}}^D \underbrace{\e^{\im \, \brackets{\mu_i - \mu_d} \frac{M - 1}{2}}}_{\text{phase term}} \, \frac{\sin \brackets{M \frac{\mu_i - \mu_d}{2}}}{\sin \brackets{\frac{\mu_i - \mu_d}{2}}} \text{,}
\end{aligned}
\label{eq:sum_analysis}
\end{equation}
where, in general, $A \brackets{\e^{\im \, \mu_i}} \neq M$, because of intermodulation products. However, if the difference $\abs{\mu_i - \mu_d}$ is equal to $\frac{2 \pi \, k}{M - 1}$, $k \in \mathbb{N}_{>0}$, the phase term is $\pm 1$, and \eqref{eq:sum_analysis} is purely real. Then, $A \brackets{\e^{\im \, \mu_i}}$ has a fixed value given by
\begin{equation}
\left. A \brackets{\e^{\im \, \mu}}\right|_{\mu_i = \mu_d \pm \frac{2 \pi \, k}{M - 1}} = M + D - 1 = \frac{\norm{\vec{a}}_2^2}{D} \text{,}
\label{eq:phase_relation}
\end{equation}
see Appendix. This can be achieved for at most
\begin{equation}
D_{\max} = M - 1
\end{equation}
different steering vectors, which corresponds to the degree of the Fourier series approximation. If the difference is smaller, there may be a noticeable phase shift from $0$ around $\mu_i$, and so the magnitude will deviate more from the value in \eqref{eq:phase_relation}. If however the difference is greater, the phase term vanishes at the sight of a relatively small value of the Dirichlet kernel as compared to the maximum of a single vector $M$. The impact factor of the phase term can be assessed by
\begin{equation}
\mathrm{IF}_i \triangleq \frac{\displaystyle \abs{A \brackets{\e^{\im \, \mu_i}}} - \frac{\norm{\vec{a}}_2^2}{D}}{\displaystyle \frac{\norm{\vec{a}}_2^2}{D}} = D \, \frac{\abs{A \brackets{\e^{\im \mu_i}}}}{\norm{\vec{a}}_2^2} - 1
\end{equation}
with $\abs{A \brackets{\e^{\im \mu_i}}} = \abs{b \brackets{\mu_i}}$, where $b \brackets{\mu_i} = \vec{a}^\CT \, \vec{a} \brackets{\mu_i} = \vec{a}^\CT \, \vec{a}_i$. It should be noted that in the case where the space sequence is a sum of steering vectors, as in \eqref{eq:vecor_sum}, $\norm{\vec{a}}_2^2 \neq D \cdot M$. Fig.~\ref{fig:phase_relation} illustrates the case where the phase relation holds. Also note that there is a mismatch between the angles of departure and the location of the two local maxima. 

\section{Root Constant-Modulus Adaptive Array}
\label{sec:root_cma_array}

In this section, we discuss in detail our proposed extension of the CMA array, which we call the root CMA array.

\subsection{Exploratory Experiment} 
\label{sec:exploratory_experiment} 

\begin{figure}[!t]
\centering
\includegraphics[height=.9\columnwidth]{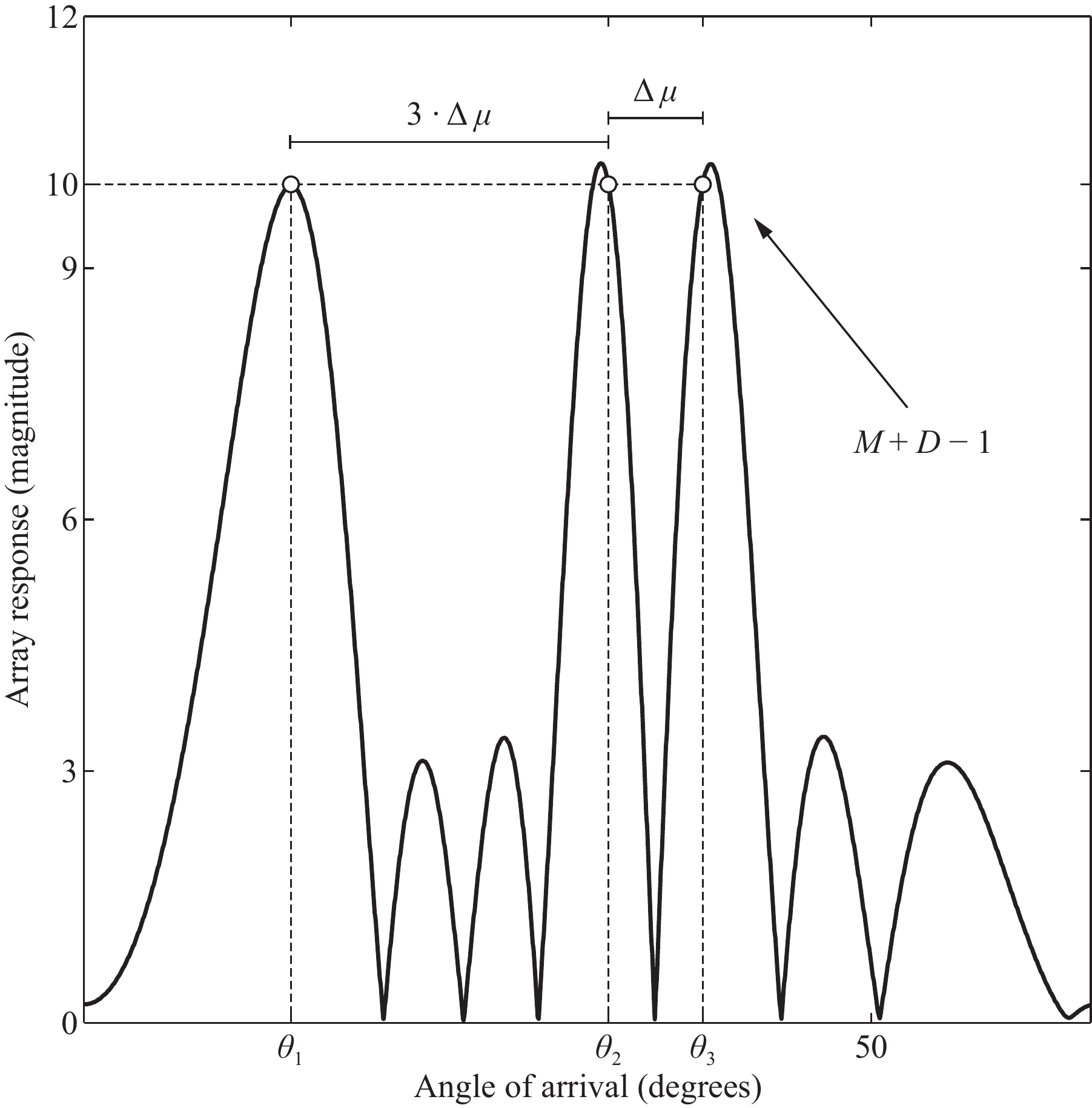}
\caption{\textcolor{red}{Inverted and normalized} magnitude response of the CMA array for $D = 3$ distinct sources for the case where the phase relation holds ($M = 8$, \textcolor{red}{$\Delta = \nicefrac{\lambda}{2}$,} $\theta_1 = -53.2^\circ$, $\theta_2 = 3.23^\circ$, $\theta_3 = 20.0^\circ$, $\SNR = 20$ dB).}
\label{fig:cma_estimate}
\end{figure}

Consider three QPSK sources being placed in the far field of a receiver in a radio system with frequency reuse. Further assume that the point sources are at equal distance, such that the incoming signals have the same amplitude. The angles of inclination of the three plane waves should obey \eqref{eq:phase_relation}, \ie,
\begin{equation}
\theta_d = \arcsin \sbrackets{\sin \theta_i \pm \frac{\lambda}{\Delta \, \brackets{M - 1}}} \text{.}
\end{equation}
No explicit orthogonality scheme such as polarization should be in use. In Fig.~\ref{fig:cma_estimate}, by way of experiment, we show that the \textcolor{red}{flipped and normalized} CMA array response is, in good approximation, equal to the sum of the steering vectors associated with the point sources (see also Fig.~\ref{fig:responses}). The array response is flipped by inverting the direction of the gradient, converting \eqref{eq:recurrence_relation} into a gradient ascent algorithm. The weight vector \textcolor{red}{$\vec{w}\brackets{n}$ is normalized by its 2-norm after each update and finally rescaled to
\begin{equation}
\vec{v}\brackets{n} = \frac{\vec{w}\brackets{n}}{\norm{\vec{w}\brackets{n}}_2} \, \sqrt{D^2 + D \, \brackets{M - 1}} \text{,}
\end{equation}
so that $\norm{\vec{v}\brackets{n}}_2 = \norm{\sum_d \vec{a}_d}_2$. This procedure is less sensitive to noise and the choice of $\gamma$ than the one mentioned in \cite{Gorlow2017_WCNPS}}. 

\subsection{Problem Restatement}

On the hypothesis that $\vec{v} \brackets{n}$ is the sum of steering vectors, the multi-user separation problem is recast as follows. Given $\vec{v} \brackets{n}$, find the set of steering vectors $\cbrackets{\hat{\vec{a}}_d}_D$ that satisfy
\begin{equation}
\vec{v}^\CT \brackets{n} \, \hat{\vec{a}} \simeq \norm{\hat{\vec{a}}}_2^2 \qquad \text{with}\ \hat{\vec{a}} = \sum_{d = 1}^D \hat{\vec{a}}_d \text{,}
\end{equation}
or, in terms of the DSFT, find
\begin{equation}
{\left\{\hat{\mu}_d \in {\left[-\pi, \pi\right)} \mid \abs{V^\ast \brackets{\e^{\im \, \hat{\mu}_d}}} \rightarrow C \right\}}
\label{eq:argmin}
\end{equation}
with $C = M + D - 1$, and $d = 1, 2, \dots, D$. In other words, we seek for the $D$ unique spatial frequencies for the steering vectors, the DSFTs of which in good approximation sum up to the DSFT of $\vec{v}\brackets{n}$. 

One can, of course, go ahead and search for the $D$ maxima of the magnitude response. However, one must be aware that not all maxima match with the underlying modes, cf.\ Fig.~\ref{fig:cma_estimate}. In the next subsection we present a more robust approach.

\subsection{Basic Approach}
\label{sec:basic_approach}

\begin{figure}[!t]
\centering
\includegraphics[height=.9\columnwidth]{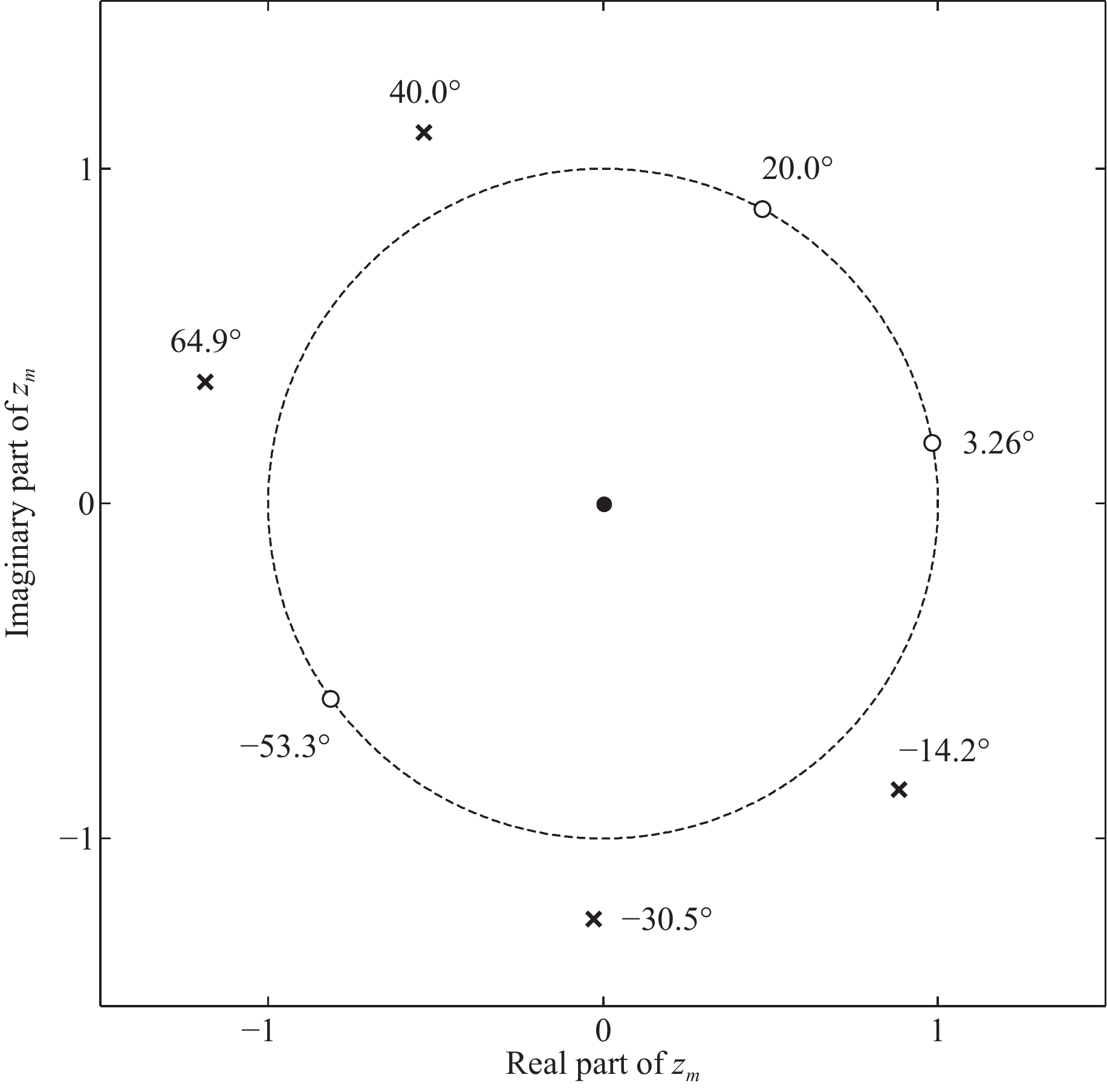}
\caption{Roots of $P \brackets{z}$ for the response shown in Fig.~\ref{fig:cma_estimate}. The solutions are marked with circles. The estimated angles of arrival are also indicated.}
\label{fig:roots}
\end{figure}

Substituting $\e^{\im \, \mu}$ for $z$, we postulate that
\begin{align}
V^\ast \brackets{z} &= v_{M - 1}^\ast \, z^{M - 1} + v_{M - 2}^\ast \, z^{M - 2} + \dots + v_0^\ast \label{eq:z_transform} \\
&\simeq C\ \brackets{= M + D - 1} \label{eq:constraint} \text{,}
\end{align}
where \eqref{eq:z_transform} is tantamount to the $z$-transform of $\vec{v} \brackets{n}$. Now if we convert \eqref{eq:constraint} into a polynomial, we obtain
\begin{equation}
P \brackets{z} = v_{M - 1}^\ast \, z^{M - 1} + \dots + \brackets{v_0^\ast - C} \text{,}
\label{eq:polynomial}
\end{equation}
which is of degree $M - 1$. For numerical stability, we further normalize \eqref{eq:polynomial} by $v_{M - 1}^\ast \neq 0$, resulting in
\begin{equation}
P \brackets{z} = \frac{v_0^\ast - C}{v_{M- 1}^\ast} + \sum_{m = 1}^{M - 1} \frac{v_m^\ast}{v_{M - 1}^\ast} \, z^m \text{.}
\label{eq:normalized}
\end{equation}
The fundamental theorem of algebra (and the factor theorem) states that $P(z)$ has $M - 1$ complex roots including possible multiplicities. Hence, the product representation of \eqref{eq:normalized} is
\begin{equation}
P \brackets{z} = \prod_{m = 1}^{M - 1} \brackets{z - z_m} \text{,}
\label{eq:product}
\end{equation}
where $z_m$ are the roots of $P\brackets{z}$. \textcolor{red}{In the noise-free case,} the sought-after frequencies are the arguments of the $D$ roots that \textcolor{red}{lie closest} to the unit circle, see Fig.~\ref{fig:roots}. The array response matrix now can be fully reconstructed with the accuracy of $\vec{v} \brackets{n}$, which (for the most part) is subject to the performance of the CMA algorithm.

\subsubsection{Analytical Solution for Two Sources}

Consider $Q\brackets{z}$ to be generated from the roots of interest only, \ie, 
\begin{equation}
Q\brackets{z} = \prod_{d = 1}^D \brackets{z - \e^{\im \, \mu_d}} \text{.}
\end{equation}
Further consider the special case where $D = 2$, then
\begin{equation}
\begin{aligned}
Q\brackets{z} &= z^2 - \brackets{\e^{\im \, \mu_1} + \e^{\im \, \mu_2}} \, z + \e^{\im \, \mu_1} \, \e^{\im \, \mu_2} \\
&= z^2 + q_1 \, z + q_0 \label{eq:quadratic_polynomial} \text{.}
\end{aligned}
\end{equation}
The roots of \eqref{eq:quadratic_polynomial} are given by the quadratic formula
\begin{equation}
z_{1, 2} = -\frac{q_1}{2} \pm \sqrt{\frac{q_1^2}{4} - q_0} \text{,}
\end{equation}
where
\begin{equation}
\begin{aligned}
q_1 &= -v_1 \brackets{n} \\
q_0 &= \Re \brackets{q_0} + \im \, \Im \brackets{q_0}
\end{aligned}
\end{equation}
with
\begin{equation}
\begin{aligned}
\Re \brackets{q_0} &= \frac{\Re^2 \sbrackets{v_1 \brackets{n}} - \Im^2 \sbrackets{v_1 \brackets{n}}}{\abs{v_1 \brackets{n}}^2} \\
\Im \brackets{q_0} &= \sin \arccos \Re \cbrackets{q_0} \text{.}
\end{aligned}
\label{eq:analytic}
\end{equation}
For a derivation of \eqref{eq:analytic} see Appendix.

\subsubsection{Numerical Solution}

Roots of polynomial equations in one unknown are numerically approximated by various known methods such as the method by Jenkins and Taub, Laguerre, or Durand and Kerner. More root-finding algorithms can be found in \cite{McNamee2007}. The function \texttt{roots} in MATLAB calculates the eigenvalues of the companion matrix, which in our case has the following structure:
\begin{equation}
{\begin{bmatrix}
-\frac{v_{M - 2}^\ast}{v_{M - 1}^\ast} & -\frac{v_{M - 3}^\ast}{v_{M - 1}^\ast} & \cdots & -\frac{v_1^\ast}{v_{M - 1}^\ast} & -\frac{v_0^\ast - C}{v_{M - 1}^\ast} \\
1 & 0 & \cdots & 0 & 0 \\
0 & 1 & \cdots & 0 & 0 \\
\vdots & \vdots & \ddots & \vdots & \vdots \\
0 & 0 & \cdots & 1 & 0
\end{bmatrix}} \text{.}
\end{equation}
The computation of eigenvalues also plays a central role in the derivation of the algebraic constant-modulus algorithm \cite{Veen1996_TSP,Veen2001_TSP,Veen2005_ICASSP}.

\section{\color{red} Adaptive CM Array Preconditioner}
\label{sec:preconditioner} 

\begin{figure*}[!t]
\centering
\includegraphics[height=.9\columnwidth]{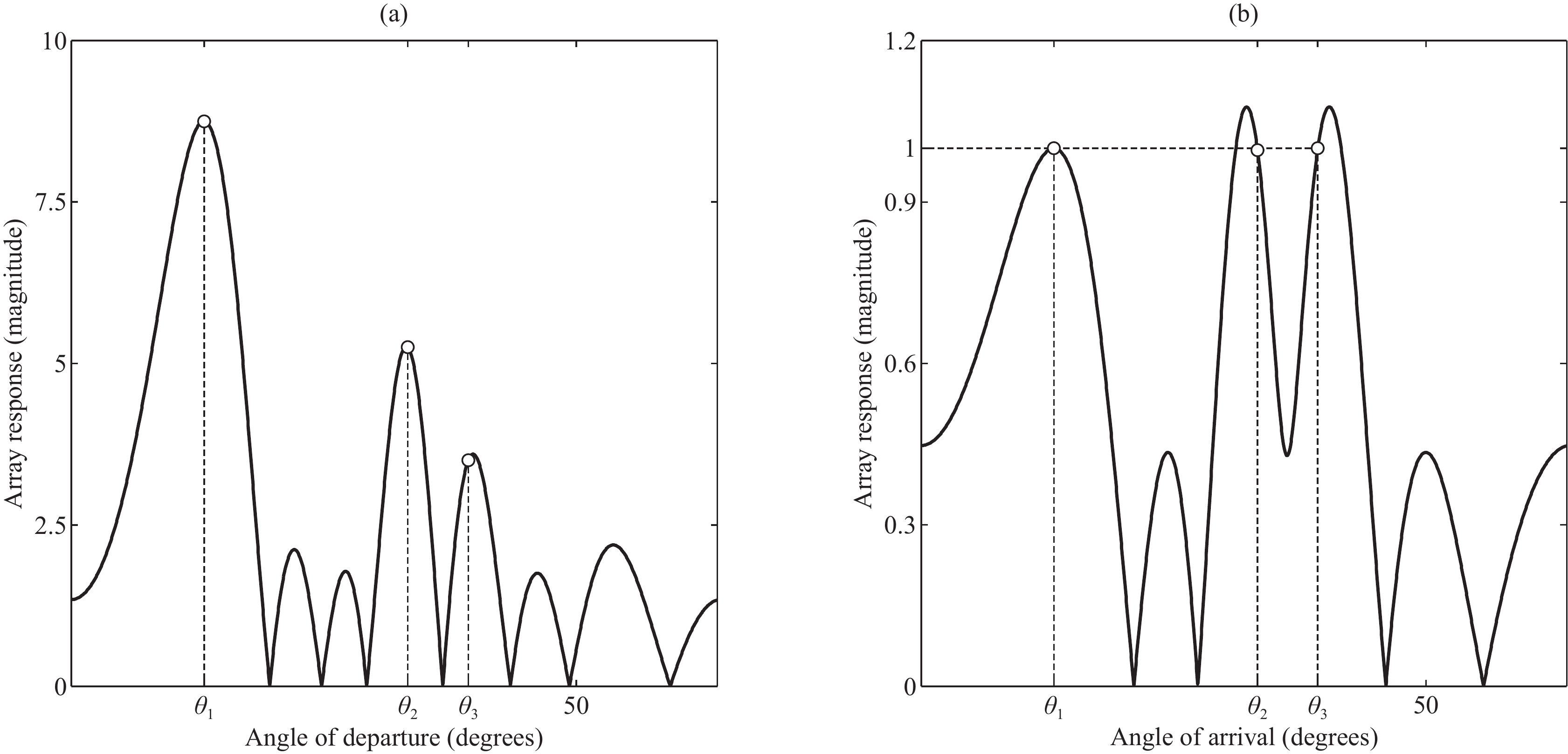}
\caption{Three QPSK sources send data from different angles at different power levels. The corresponding radiation pattern is shown in (a). The response of the preprocessor after 1,000 iterations is shown in (b). The array consists of 8 elements \textcolor{red}{with half-wavelength spacing}. The angles of interest correspond to those in Fig.~\ref{fig:cma_estimate}.}
\label{fig:preprocessor}
\end{figure*}

As previously mentioned, the central problem of the CMA algorithm is its inability to lock onto one single user without proper initialization of the weight vector or preprocessing. A possible solution to this problem is illustrated in the previous subsection. Nonetheless, there are two main limitations linked to that solution. Firstly, the number of users (model order) is assumed to be known a priori. Secondly, the incoming waves are supposed to have the same amplitude. In this section, we present an alternative and more practical solution in the form of an adaptive preprocessor for the CMA array. It tackles the model-order estimation problem, plus it allows the signals to differ in amplitude.

\color{red}
\subsection{Motivation}

The empirical evidence, which underlies our original work in \cite{Gorlow2017_WCNPS} and which supports the hypothesis that the response of the CMA array (if inverted) is proportional to the sum of the array steering vectors when the array is initialized, e.g., with an all-pass response, can be interpreted in the following way: After the first iteration, the output is equal to the sum of the constant modulus signals. With every new iteration the array is adjusted in such a way as to minimize the cost, which is a function of the squared magnitude of $y(n)$, 
\begin{equation}
\E{\cbrackets{\abs{y(n)}^2}} \propto \vec{w}^\CT(n) \, \vec{A} \, \vec{A}^\CT \, \vec{w}(n) \quad \text{with} \ \cbrackets{s_i(n)} \  \text{\emph{i.i.d.}} \text{.} 
\label{eq:squared_magnitude}
\end{equation}
In general, \eqref{eq:squared_magnitude} is positive and real for any $\vec{w}(n) \neq \vec{0}$, which includes 
\begin{equation}
\vec{w} = \sum_{d = 1}^D \vec{a}_d \quad \text{with} \ w_0 = D 
\end{equation}
as some sort of a (spatially) matched filter for the transmitted data. The corresponding polynomial is 
\begin{equation}
P\brackets{z} = w_{M - 1}^\ast \, z^{M - 1} + \dots +1 - M \text{.} 
\label{eq:matched_polynomial}
\end{equation}
Since $M > 1$, \eqref{eq:matched_polynomial} resembles the transfer function of a linear prediction error filter for a scaled sum of $D$ symbols. Insight and intuition tells us that in a similar manner one can design a more performant predictor that also exploits the phase, see \cite{Tufts1982}. 

\subsection{Ordinary Least Squares}

Ordinary least squares (OLS) is the most common and the most basic estimator that can be employed to carry out linear prediction on the CM array. Extensions of the former with or without additional modifications are discussed and evaluated in \cite{Tufts1982} and later works, such as \cite{Dowling1996, Lopes2002}. 

Consider the desired or predicted signal to be equal to the signal at the first antenna element, and allow the CM signals to differ in amplitude $c$, i.e. 
\begin{equation}
x_0\brackets{n} = \sum_{d = 1}^D{c_d \, s_d\brackets{n}} \text{.}
\label{eq:desired_signal} 
\end{equation}
The optimum filter coefficients in the least squares sense are obtained by minimizing the sum of squared residuals, which are defined as the differences between $x_0(n)$ and $y(n)$, where $y(n)$ is predicted from the remaining $M - 1$ array signals. In consequence, the OLS solution yields 
\begin{equation}
\vec{w} = \rbrackets{\vec{X}_{1,\dots,M-1} \, \vec{X}_{1,\dots,M-1}^\CT}^{-1} \, \vec{X}_{1,\dots,M-1} \, \vec{x}_0^\CT \in \C^{M - 1} \text{,} 
\end{equation}
where 
\begin{equation}
\vec{x}_0 = \begin{bmatrix} x_0(0) & x_0(1) & \cdots & x_0(N - 1) \end{bmatrix} \in \C^{1 \times N} 
\end{equation}
and
\begin{equation}
\vec{X}_{1,\dots,M-1} = \begin{bmatrix} \vec{x}_1 \\ \vec{x}_2 \\ \vdots \\ \vec{x}_{M - 1} \end{bmatrix} \in \C^{M - 1 \times N} \text{,} 
\end{equation}
respectively.

\color{black}

\subsection{Adaptive Preprocessor}

\begin{figure}[!t]
\centering
\includegraphics[height=.9\columnwidth]{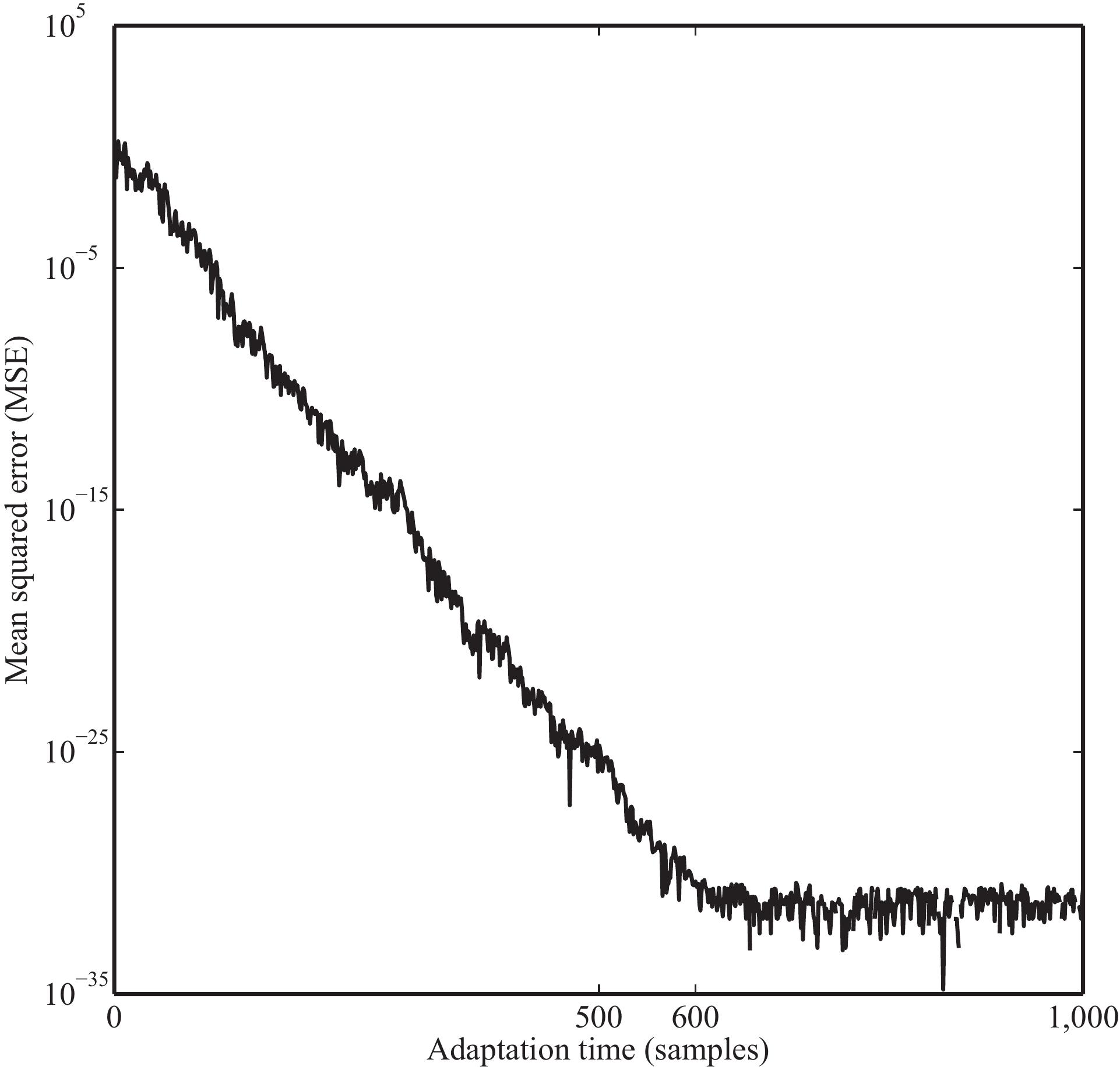}
\caption{Learning curve of the preprocessor in the noise-free case with $\gamma_n = \nicefrac{2}{\norm{\vec{x}\brackets{n}}_2^2} - \epsilon$. The saddle point is reached after 600 iterations.}
\label{fig:learning}
\end{figure}

\textcolor{red}{In reference to OLS,} the \textcolor{red}{adaptive} preprocessor is designed as an LMS filter with the desired signal \textcolor{red}{as in \eqref{eq:desired_signal}}. The LMS filter \textcolor{red}{is initialized} with an all-zero response \textcolor{red}{$\vec{u} = \vec{0}_{M \times 1}$}. Then, only the last $M - 1$ weights are updated \textcolor{red}{by} 
\textcolor{red}{\begin{equation}
\vec{u}\brackets{n + 1} = {\begin{bmatrix} 0 \\
u_1\brackets{n} + \gamma \, x_1\brackets{n} \, e^\ast\brackets{n} \\
\vdots \\
u_{M - 1}\brackets{n} + \gamma \, x_{M - 1}\brackets{n} \, e^\ast\brackets{n} 
\end{bmatrix}} \text{,}
\end{equation}}%
where
\begin{equation}
e\brackets{n} = x_0\brackets{n} - y\brackets{n}
\end{equation}
and
\begin{equation}
y\brackets{n} = \vec{u}^\CT\brackets{n} \, \vec{x}\brackets{n} \text{.}
\end{equation}
In this way, we avoid that the LMS filter converges to an all-pass response, which would be the optimum solution but not exactly what we are after. Instead, we force the filter to give us an LMS estimate of $x_0\brackets{n}$ using the delayed versions. The hypothesis is that the filter will adjust its coefficients in such a way that separate array responses point in the directions of the users. The step size can be chosen adaptively fulfilling
\begin{equation}
\gamma_n < \frac{2}{\norm{\vec{x}\brackets{n}}_2^2} \text{,}
\label{eq:gamma_condition}
\end{equation}
which keeps the filter stable. Fig.~\ref{fig:preprocessor} shows the filter response after 1,000 iterations alongside the radiation pattern. One can observe that the filter response takes the value of one exactly where the modes are. That was to be expected, because $x_0\brackets{n}$ and $y\brackets{n}$ share the same signal amplitudes $\cbrackets{c_d}$. Thus, we can use the same approach as before to determine the modes. In Fig.~\ref{fig:learning} we see the learning curve of the preprocessor for the noise-free case. The algorithm converges after 600 iterations \textcolor{red}{for the largest possible $\gamma$ according to \eqref{eq:gamma_condition}}. The mean squared error is virtually zero, \ie, $x_0\brackets{n}$ and $y\brackets{n}$ are almost identical.  

\subsection{Model-Order \textcolor{red}{Selection}}

\begin{figure}[!t]
\centering
\includegraphics[height=.9\columnwidth]{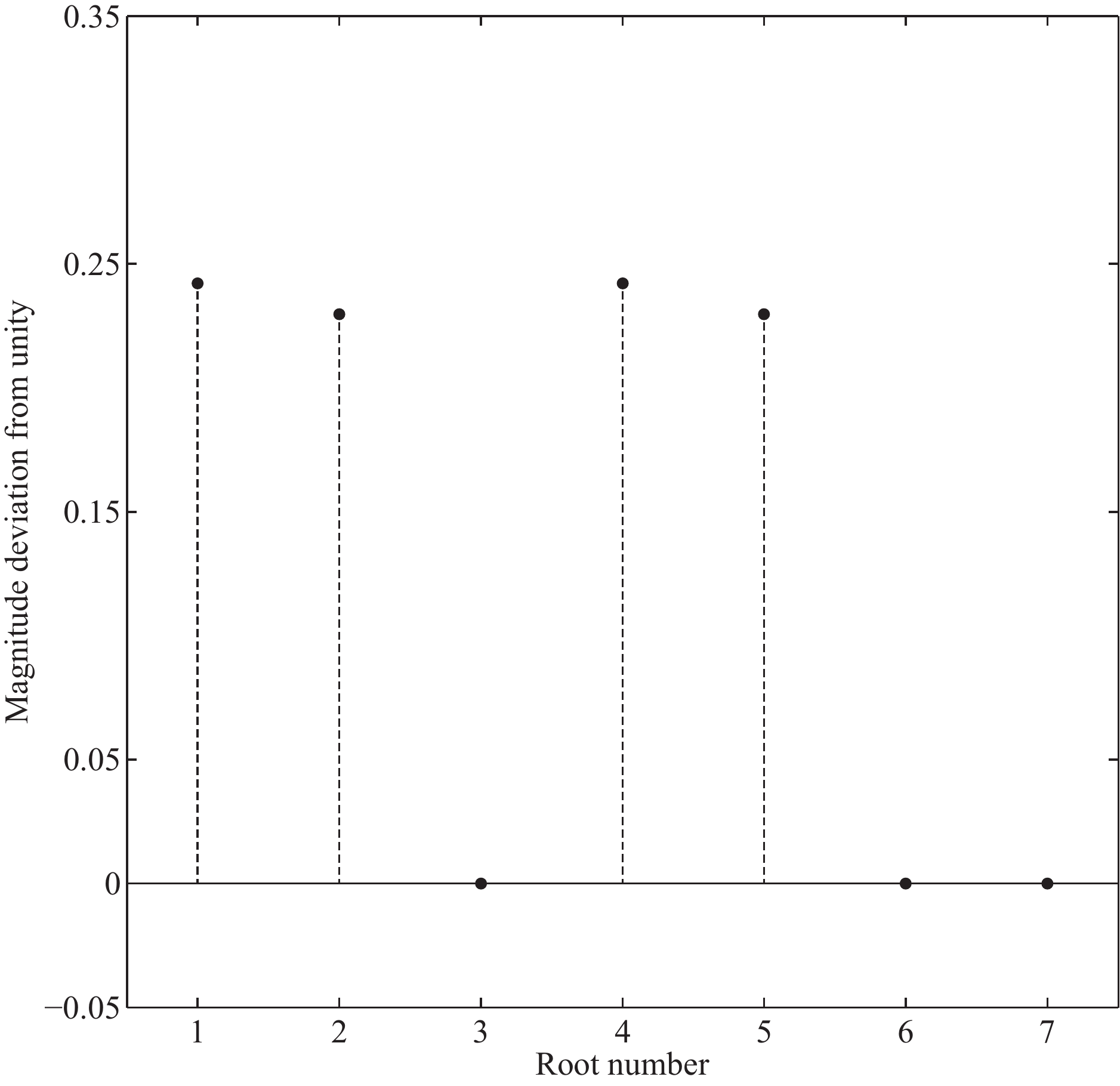}
\caption{Deviation of the polynomial roots from unity (noise-free case). As can be seen, three roots lie exactly on the unit circle \textcolor{red}{($D = 3$)}.}
\label{fig:deviation}
\end{figure}

The model order is given by the number of roots that \textcolor{red}{lie closest} to the unit circle, see Fig.~\ref{fig:roots}. We may consider, \eg, the distance $\abs{z_m} - 1$ to detect that number. This is illustrated in Fig.~\ref{fig:deviation}. Note that in order to compute the roots correctly, $C$ in \eqref{eq:normalized} must be set to $1$. As a more sophisticated alternative, and especially when the data is noisy, the Neyman--Pearson lemma can prove useful \cite{Silverstein1994}. A comparison of more and other schemes is made in \cite{DaCosta2009}. 

\textcolor{red}{In our experiments, however, we found that when the noise level is high and the spatial frequencies are chosen arbitrarily, it is more reliable to evaluate the real part of the array response at the inclination angles that correspond to the $M - 1$ roots (see next section) than to evaluate the distance of the roots from the unit circle. We therefore suggest to consider the $D$ roots with the strongest (real) array response as the signal roots.}

\subsection{Direction-of-Arrival Estimation}

If the model order $D$ is determined, one can easily derive the directions of arrival from the phase of the corresponding roots. The inclination angles then are given by
\begin{equation}
\theta_d = \arcsin\brackets{\frac{\lambda \, \arg z_d}{2 \pi \, \Delta}} \text{,}
\end{equation}
for $d = 1, 2, \dots, D$.

\subsection{Array Preconditioning}

With the model order $D$ known and the directions of arrival at hand, one can reconstruct the array response matrix $\vec{A}$. The pseudoinverse of $\vec{A}$, $\vec{A}^+$, has the property that the main lobe of each steering vector points in the direction of the respective source. All other sources, \ie, the interferers, are fully suppressed. Hence, one can initialize the weights of the CMA array with the row vector of $\vec{A}^+$ that corresponds to the source \textcolor{red}{(user)} of interest and subsequently run the CMA algorithm. \textcolor{red}{The analysis of the combined performance of the preconditioner and the CMA array is beyond the scope of this article.} 

\subsection{Summary}

In summary, the \textcolor{red}{preconditioned} CMA algorithm consists of the steps that are listed below:
\begin{enumerate}
	\item Initialize \textcolor{red}{LMS} filter with all-zero response;
	\item \textcolor{red}{Consider signal at first element as desired response;} 
	\item Update filter weights using gradient descent;
	\item Exclude all-pass response;
	\item Compute roots of corresponding polynomial;
	\item Estimate model order;
	\item Choose roots with \textcolor{red}{strongest beam response};
	\item Reconstruct array response matrix;
	\item Run classical CMA algorithm (optional).
\end{enumerate}
The individual steps are elaborated further in Fig.~\ref{fig:preproc} in the form of pseudo code.

\begin{figure}[!t]
\small
\begin{framed}
\begin{spacing}{1.1}
\begin{algorithmic}[1]
\Procedure{rootcma}{$\vec{X}, \gamma$}
	\State $\vec{u} \gets \sbrackets{\begin{smallmatrix} 0 & 0 & \cdots & 0 \end{smallmatrix}}^\T$ \Comment{all-zero response}
	\State \textcolor{red}{$\bar{\vec{u}} \gets \sbrackets{\begin{smallmatrix} 1 & 0 & \cdots & 0 \end{smallmatrix}}^\T$ \Comment{desired response}}
	\For{\textcolor{red}{$n \gets 1, 2, \dots, N$}}
		\State \textcolor{red}{$\vec{x} \gets \sbrackets{\begin{smallmatrix} x_{\ast, n} \end{smallmatrix}}$} \Comment{$M$ snapshots}
		\State \textcolor{red}{$y \gets \vec{u}^\CT \, \vec{x}$} \Comment{estimate}
		\State \textcolor{red}{$e \gets x_1 - y$} \Comment{\textcolor{red}{error}}
		\State $\vec{u} \gets \vec{u} + \gamma \, \vec{x} \, e^\ast$ \Comment{update}
		\State \textcolor{red}{$u_1 \gets 0$} \Comment{exclude all-pass response}
		\State \textcolor{red}{$\vec{p} \gets$ \Call{flipud}{$\vec{u}^\ast - \bar{\vec{u}}$} \Comment{prediction error polynomial}}
		\State \textcolor{red}{$\vec{z} \gets$ \Call{roots}{$\vec{p}$}} \Comment{polynomial roots}
		{\color{red}\For{$d \gets 1, 2, \dots, M - 1$}
			\For{$m \gets 0, 1, \dots, M - 1$}
				\State $a_{md} \gets \exp\brackets{\im \arg z_d \, m}$ \Comment{response matrix}
			\EndFor
		\EndFor
		\State $\vec{b} \gets$ \Call{real}{$\vec{u}^\CT \, \vec{A}$} \Comment{beam response}
		\State $\vec{b} \gets$ \Call{sort}{$\vec{b}, \vec{i}$} \Comment{sort in descending order}
		\State $D \gets$ \Call{find}{$\vec{b} > \epsilon, 1$}${} - 1$ \Comment{model order}
		\State $\vec{A} \gets \sbrackets{\begin{smallmatrix} a_{\ast, \sbrackets{\begin{smallmatrix} i_1, i_2, \dots, i_D \end{smallmatrix}}} \end{smallmatrix}}$ \Comment{strongest response matrix}
		\State $\vec{W} \gets \vec{A} \, \brackets{\vec{A}^\T \, \vec{A}}^{-1}$ \Comment{beam steering matrix}}
		\State $\ldots$ \Comment{CMA array, etc.}
	\EndFor
\EndProcedure
\end{algorithmic}
\end{spacing}
\end{framed}
\caption{Root constant-modulus adaptive algorithm (pseudo code).}
\label{fig:preproc}
\end{figure}

\section{Conclusion}
\label{sec:conclusion}

We presented a novel approach for the separation of multiple sources in a frequency-reuse radio system making use of the classical CMA array. Our extension of the CMA array makes it applicable when one particular source is to be captured in the presence of multiple CM sources. The cited literature relies on an implicit capture effect of the algorithm, which is hardly sustainable looking at the formulation of the underlying cost function. The CMA array rather acts as a soft spatial equalizer. This fact can be exploited to reconstruct the array response matrix and to find the best fit for the constant modulus signals in, e.g., the least-squares sense immediately after. In addition, we designed an alternative preprocessor for the array based on the LMS filter. The second solution is more practical, because it allows the CM signals to differ in amplitude. 

\appendix

\section*{Derivation of Equation \eqref{eq:phase_relation}}

In any such case where the phase relation $\abs{\mu_i - \mu_d} = \frac{2 \pi \, k}{M - 1}$ holds, \eqref{eq:sum_analysis} becomes
\begin{align}
A \brackets{\e^{\im \, \mu_i}} &= M + \sum_{\substack{d = 1\\ d \neq i}}^D \e^{\im \, \pi k} \, \frac{\sin \brackets{M \frac{\pi k}{M - 1}}}{\sin \brackets{\frac{\pi k}{M - 1}}} \nonumber \\
&= M + \sum_{\substack{d = 1 \\ d \neq i}}^D \brackets{-1}^k \, \frac{\sin \brackets{\frac{\pi k}{M - 1} + \pi k}}{\sin \brackets{\frac{\pi k}{M - 1}}} \nonumber \\
&= M + \sum_{\substack{d = 1 \\ d \neq i}}^D \brackets{-1}^k \, \brackets{-1}^k \nonumber \\
&= M + D - 1 \qquad \forall \, M, k \in \mathbb{N}_{>0}\text{.} \label{eq:root_value}
\end{align}
Furthermore, if we consider that
\begin{align}
\norm{\vec{a}}_2^2 &= D^2 + \sum_{m = 1}^{M - 1} a_m^\ast \, a_m \qquad \vline \quad a_m = \sum_{d = 1}^D \e^{\im \, \mu_d m}  \nonumber \\
&= D^2 + \sum_{m = 1}^{M - 1} \sum_{d = 1}^D \e^{-\im \, \mu_d m} \sum_{i = 1}^D \e^{\im \, \mu_i m} \nonumber \\
&= D^2 + \sum_{m = 1}^{M - 1} \sum_{i = 1}^D \sum_{d = 1}^D \e^{\im \, \brackets{\mu_i - \mu_d} \, m} \nonumber \\
&= D^2 + \sum_{m = 1}^{M - 1} \sum_{i = 1}^D \sum_{d = i}^D \sbrackets{\e^{\im \, \brackets{\mu_i - \mu_d} \, m} + \e^{-\im \, \brackets{\mu_i - \mu_d} \, m}} \nonumber \\
&= D^2 + \sum_{m = 1}^{M - 1} \sbrackets{D + 2 \sum_{i = 1}^D \sum_{d = i + 1}^D \Re \cbrackets{\e^{\im \, \brackets{\mu_i - \mu_d} \, m}}} \text{,} 
\label{eq:norm_unfolded}
\end{align}
and if we use the phase relation from above, \eqref{eq:norm_unfolded} writes
\begin{align}
\norm{\vec{a}}_2^2 &= D^2 + D \, \brackets{M - 1} + 2 \underbrace{\sum_{m = 1}^{M - 1} \cos \brackets{\frac{2 \pi \, m k}{M - 1}}}_{= \: 0 \ \forall \, M, \, k \in \, \mathbb{N}_{>0}} \nonumber \\
&= D^2 + D \, \brackets{M - 1} \qquad \forall \, M, k \in \mathbb{N}_{>0} \text{.} \label{eq:norm_special}
\end{align}
Comparing \eqref{eq:root_value} with \eqref{eq:norm_special}, we see that
\begin{equation}
A \brackets{\e^{\im \, \mu_i}} = M + D - 1 = \frac{\norm{\vec{a}}_2^2}{D} \text{.} \qquad \blacksquare
\end{equation}

\section*{Derivation of Equation \eqref{eq:analytic}}
\label{app:analytical}

For $D = 2$ and for a given $\vec{v} \brackets{n}$, we know that
\begin{equation}
\begin{aligned}
\e^{\im \, \mu_1} + \e^{\im \, \mu_2} &= \brackets{\cos \mu_1 + \cos \mu_2} + \im \, \brackets{\sin \mu_1 + \sin \mu_2} \\
&= \Re \sbrackets{v_1 \brackets{n}} + \Im \sbrackets{v_1 \brackets{n}} \text{.}
\end{aligned}
\label{eq:known}
\end{equation}
What we seek for is
\begin{equation}
\begin{aligned}
\e^{\im \, \mu_1} \, \e^{\im \, \mu_2} &= \cos \brackets{\mu_1 + \mu_2} + \im \sin \brackets{\mu_1 + \mu_2} \\
&= \Re \brackets{q_0} + \Im \brackets{q_0} \text{.}
\end{aligned}
\end{equation}
Resorting to trigonometric identities, we find that
\begin{align}
&\cos \mu_1 + \cos \mu_2 \nonumber \\
&\qquad = 2 \cos \frac{\mu_1 + \mu_2}{2} \, \cos \frac{\mu_1 - \mu_2}{2} \nonumber \\
&\qquad = \sqrt{\sbrackets{1 + \cos \brackets{\mu_1 + \mu_2}} \, \sbrackets{1 + \cos \brackets{\mu_1 - \mu_2}}} \label{eq:cos}
\end{align}
and
\begin{align}
&\sin \mu_1 + \sin \mu_2 \nonumber \\
&\qquad = 2 \sin \frac{\mu_1 + \mu_2}{2} \, \cos \frac{\mu_1 - \mu_2}{2} \nonumber \\
&\qquad = \sqrt{\sbrackets{1 - \cos \brackets{\mu_1 + \mu_2}} \, \sbrackets{1 + \cos \brackets{\mu_1 - \mu_2}}} \label{eq:sin} \text{,}
\end{align}
respectively. Dividing \eqref{eq:cos} by \eqref{eq:sin}, we obtain
\begin{equation}
\frac{\cos \mu_1 + \cos \mu_2}{\sin \mu_1 + \sin \mu_2} = \sqrt{\frac{1 + \cos \brackets{\mu_1 + \mu_2}}{1 - \cos \brackets{\mu_1 + \mu_2}}} \label{eq:cos_by_sin} \text{.}
\end{equation}
From \eqref{eq:cos_by_sin} it follows that
\begin{align}
&\cos \brackets{\mu_1 + \mu_2} \nonumber \\
&\qquad = \frac{\brackets{\cos \mu_1 + \cos \mu_2}^2 - \brackets{\sin \mu_1 + \sin \mu_2}^2}{\brackets{\cos \mu_1 + \cos \mu_2}^2 + \brackets{\sin \mu_1 + \sin \mu_2}^2} \label{eq:cos_sum} \text{.}
\end{align}
By comparing the right-hand side of \eqref{eq:known} with the right-hand side of \eqref{eq:cos_sum}, we immediately see that
\begin{equation}
\cos \brackets{\mu_1 + \mu_2} = \frac{\Re^2 \sbrackets{v_1 \brackets{n}} - \Im^2 \sbrackets{v_1 \brackets{n}}}{\abs{v_1 \brackets{n}}^2} \label{eq:cos_sum_known} \text{.} \qquad \blacksquare
\end{equation}

\section*{Acknowledgment}

The authors would like to thank the Brazilian research and innovation agencies FAPDF (Funda\c{c}\~ao de Apoio \`a Pesquisa do Distrito Federal), CAPES (Coordena\c{c}\~ao de Aperfei\c{c}oamento de Pessoal de N\'ivel Superior), and CNPq (Conselho Nacional de Desenvolvimento Cient\'ifico e Tecnol\'ogico).

\nocite{Papadias1997}
\nocite{Xu2001}

\bibliographystyle{IEEEtran}
\bibliography{IEEEabrv,references,wcnps2017}

\end{document}